\documentclass[11pt,preprint]{aastex}
\usepackage{amssymb}
\usepackage{epsfig}
\usepackage{natbib}
\usepackage{graphicx}
\usepackage{color}
\usepackage{tabularx}

\newcommand       \mum        {\,{\rm \mu m}}
\newcommand       \V            {{\rm V}}

\newcommand       \Ks           {{\rm K_{s}}}
\newcommand       \J            {{\rm J}}
\newcommand       \JJ            {{\rm J}}
\newcommand       \HH           {{\rm H}}

\newcommand       \K            {\,{\rm K}}
\newcommand       \simali       {\,{\sim}}
\newcommand       \magni        {\,{\rm mag}}
\newcommand       \simlt        {\lesssim}
\newcommand       \simgt        {\gtrsim}
\newcommand       \gtsim        {\gtrsim}

\newcommand       \Angstrom     {\,{\rm \AA}}

\newcommand       \cm           {\,{\rm cm}}

\newcommand       \km           {\,{\rm km}}

\newcommand       \s            {\,{\rm s}}

\newcommand       \nH           {n_{\rm H}}

\newcommand       \Tgas       {T_{\rm gas}}
\newcommand       \amin       {a_{\rm min}}
\newcommand       \amax       {a_{\rm max}}

\newcommand{\AV}{A_{\rm V}}

\newcommand{\RV}{R_{\rm V}}
\newcommand{\Rv}{R_{\rm V}}

\countdef\decade=200
\advance\decade by \year
\countdef\hours=201
\advance\hours by \time
\divide\hours by 60
\countdef\mins=202
\advance\mins by \hours
\multiply\mins by 60
\multiply\hours by 100
\countdef\miltime=203
\advance\miltime by \hours
\advance\miltime by \time
\advance\miltime by -\mins

\shorttitle{The Mid-Infrared Extinction Law in the Coalsack Nebula}
\shortauthors{Wang et al.}

\received{2013 06 10}
\begin{document}
\title{
The Mid-Infrared Extinction Law
and its Variation in the Coalsack Nebula
     }
\author{Shu Wang\altaffilmark{1,2},
             Jian Gao\altaffilmark{1},
             B.W.~Jiang\altaffilmark{1},
             Aigen Li\altaffilmark{2}, and
             Yang Chen\altaffilmark{1}}
\altaffiltext{1}{Department of Astronomy,
                 Beijing Normal University,
                 Beijing 100875, China;
                 {\sf shuwang@mail.bnu.edu.cn,
                      jiangao@bnu.edu.cn,
                      bjiang@bnu.edu.cn, cheny@bnu.edu.cn
                 }
                 }
\altaffiltext{2}{Department of Physics and Astronomy,
                 University of Missouri,
                 Columbia, MO 65211, USA;
                 {\sf lia@missouri.edu}
                 }

\begin{abstract}
In recent years the wavelength dependence of interstellar extinction
from the ultraviolet (UV), optical, through the near-
and mid-infrared (IR) has been studied extensively.
Although it is well established that the UV/optical extinction law
varies significantly among the different lines of sight, it is not
clear how the IR extinction varies among various environments.
In this work, using the color-excess method and taking red giants
as the extinction tracer, we determine the interstellar extinction
$A_\lambda$ in the four
{\it Spitzer}/IRAC bands in [3.6], [4.5], [5.8], [8.0]$\mum$
(relative to $A_\Ks$, the extinction in
the {\it 2MASS} $\Ks$ band at 2.16$\mum$)
of the Coalsack nebula, a nearby starless dark cloud,
based on the data obtained from the {\it 2MASS}
and \textit{Spitzer}/GLIMPSE surveys.
We select five individual regions across the nebula that span a wide
variety of physical conditions, ranging from diffuse, translucent to dense
environments, as traced by the visual extinction,
the {\it Spitzer}/MIPS 24$\mum$ emission, and CO emission.
We find that $A_\lambda/A_\Ks$,
the mid-IR extinction relative to $A_\Ks$,
decreases from diffuse to dense environments, which may be
explained in terms of ineffective dust growth in dense regions.
The mean extinction (relative to $A_\Ks$) is calculated for
the four IRAC bands as well, which exhibits a flat mid-IR
extinction law, consistent with previous determinations
for other regions.
The extinction in the IRAC 4.5$\mum$ band is anomalously
high, much higher than that of the other three IRAC bands.
It cannot be explained in terms of CO and CO$_2$ ices.
The mid-IR extinction in the four IRAC bands have also
been derived for four representative regions
in the Coalsack Globule 2
which respectively exhibit strong ice absorption,
moderate or weak ice absorption,
and very weak or no ice absorption.
The derived mid-IR extinction curves are all flat,
with $A_{\lambda}/A_\Ks$ increasing
with the decrease of the 3.1$\mum$ H$_2$O ice absorption
optical depth $\tau_{\rm ice}$.
\end{abstract}


\section{Introduction}
\subsection{Diffuse, Translucent, and Dense Clouds}
The interstellar medium (ISM) is generally classified into three
phases (see Snow \& McCall 2006):
the cold neutral medium (CNM), the warm ionized medium or
warm neutral medium (WIM or WNM), and the hot ionized medium
(HIM; also known as the intercloud medium or the coronal gas).
The CNM itself contains a variety of cloud types
(e.g., diffuse clouds, translucent clouds and molecular clouds),
spanning a wide range of physical and chemical conditions.
In diffuse clouds
(e.g., sightlines toward $\zeta$ Oph and $\zeta$ Per)
hydrogen is mostly in atomic form and carbon is mostly
in ionized form (C$^{+}$). They typically have a total
visual extinction $\AV\sim 0-1\magni$, a hydrogen number
density $\nH\sim 10-500\cm^{-3}$, and a gas  temperature
$\Tgas \sim 30-100\K$.\footnote{%
   Snow \& McCall (2006) further classified diffuse clouds into
   diffuse atomic clouds and diffuse molecular clouds. For the latter,
    the fraction of H in molecular form becomes appreciable ($>$\,10\%),
    but C is still predominantly in the form of C$^{+}$.
    Van Dishoeck (1994) defined diffuse clouds as regions in which
    the transition of H from atomic to molecular form occurs.
    \label{ft:cloud}
    }
In translucent clouds, hydrogen is mostly in molecular (H$_2$)
and the transition of carbon from ionized (C$^{+}$) to atomic (C)
or molecular (CO) form takes place (e.g., the high latitude cirrus
clouds, see Magnani et al.\  1985).
These clouds have $\AV\sim1-5\magni$,
$\nH\sim 500-5000\cm^{-3}$, and $\Tgas \sim 15-50\K$
and are thick enough for CO millimeter lines to be detectable.
In molecular clouds (which are variously referred to as dense clouds,
dark clouds, or dark molecular clouds)\footnote{%
   In this work we distinguish two types of molecular clouds:
   dark clouds with $5< \AV<20\magni$
   and infrared (IR) dark clouds with $\AV>20\magni$.
   }
carbon becomes almost completely molecular (CO).
They are cold (with $\Tgas\sim 10-15\K$),
subject to large dust extinction (with $\AV > 5-10\magni$),
 and their densities typically exceed $10^4\cm^{-3}$.
We note that dense cloud material is often surrounded by
translucent material which, in turn, is often surrounded
by diffuse cloud material.

Interstellar clouds become either diffuse, translucent or molecular,
depending on the extent to which ultraviolet (UV) photons can
penetrate into the clouds: with little extinction, diffuse clouds are
almost fully exposed to the interstellar UV radiation so that nearly
all molecules are destroyed by photodissociation
(but also see Footnote \ref{ft:cloud}),
in molecular clouds where the large dust extinction highly
attenuates the interstellar UV radiation, molecules are protected
from being photodissociated, while translucent clouds fall in between
these two extremes.

Dust plays a crucial role in governing the physical and chemical
states of interstellar molecules and determining the physical and chemical
structures of interstellar clouds. Apart from providing surfaces for
the formation of some molecules (e.g., H$_2$), dust, through the
scattering and absorption of starlight, controls the attenuation of
UV radiation passing through the clouds and the depth to
which UV radiation penetrates into the clouds.
Hence dust extinction -- the sum of absorption and scattering --
is a crucial parameter in modeling the chemistry of interstellar clouds.

\subsection{UV/Optical Extinction:
                    Strong Dependence on Environments}
The interstellar extinction law -- the variation of
extinction $A_\lambda$ with wavelength $\lambda$,
usually expressed as $A_\lambda/A_{\rm V}$
-- is known to vary significantly among sightlines
in the UV/optical at $\lambda<0.7\mum$
(see Cardelli et al.\ 1989).\footnote{%
    This means of expressing the extinction law is not unique;
    it has also been common practice to use instead the ratios of
    two colors, $E(\lambda-V)/E(B-V)$,
    where $E(\lambda-V)\equiv A_\lambda - A_{\rm V}$,
    $A_{\rm B}$ is the extinction in the blue ($B$) band,
    and $A_{\rm V}$ is the extinction in the visual ($V$) band.
    }
\citet{Cardelli89} found that the variation in the UV/optical can be
described by a functional form (hereafter ``CCM'')
with only one parameter, $R_{\rm V}\,\equiv\,A_{\rm V}/E(B-V)$,
the optical total-to-selective extinction ratio.
The value of $R_{\rm V}$ depends upon the environment along
the line of sight. A direction through diffuse, low-density regions
usually has a rather low value of $R_{\rm V}$.
The ``average" extinction law for the Galactic diffuse clouds
is described by a CCM extinction curve with $R_{\rm V} \approx 3.1$,
which is commonly used to correct observations for dust extinction.
Lines of sight penetrating into dense clouds,
such as the Ophiuchus or Taurus molecular clouds,
usually show $4 < R_{\rm V} < 6$ (see Mathis 1990).\footnote{%
  In the literature, the UV/optical extinction laws have mainly been
  measured for diffuse clouds, translucent clouds, and the outskirts of
  molecular clouds. For sources deeply embedded in dense clouds
  only the IR extinction can be obtained.
  }
Theoretically, $R_{\rm V}$ may become infinity in dense regions rich
in very large, ``gray'' grains (i.e., the extinction caused by these
grains does not vary much with wavelength),
while the steep extinction produced completely by Rayleigh scattering
would have $R_{\rm V}\sim 1.2$ \citep{Draine03}.
A wide range of $R_{\rm V}$ values have been reported for extragalactic
lines of sight, ranging from $R_\V\approx 0.7$ for a quasar
intervening system at $z\approx1.4$ (Wang et al.\ 2004)
to $R_\V\approx 7$ for gravitational lensing galaxies (Falco et al.\ 1999).

\subsection{Near-IR Extinction: A Universal Power Law?}
With the wealth of available data from space-borne telescopes
(e.g., ISO and Spitzer) and ground-based surveys (e.g., 2MASS)
in the near- and mid-IR, in recent years we have seen an increase
in interest in IR extinction. Understanding the effects of
dust extinction in the IR wavelengths is important to properly interpret
these observations.

As the UV/optical extinction varies substantially among various
environments, one might expect the near-IR extinction
at $0.9\mum < \lambda <3\mum$ to vary correspondingly.
However, for over two decades astronomers have believed that there is
little, if any, near-IR extinction variation
from one line of sight to another.
The near-IR extinction law appears to be an approximately uniform
power law of $A_\lambda \sim \lambda^{-\alpha}$
with $\alpha$\,$\approx$\,1.6--1.8,
independent of environment or $R_{\rm V}$
at $0.9\mum < \lambda < 3\mum$.\footnote{%
   Martin \& Whittet (1990) found $\alpha\approx1.8$
   in the diffuse ISM as well as the outer regions of
   the $\rho$ Oph and Tr\ 14/16 clouds.
   Using a large sample of obscured OB stars, He et al.\ (1995)
   found that the near-IR extinction can be well-fitted to a power law
   with $\alpha\approx 1.73\pm0.04$, even though the $R_{\rm V}$ value
   may vary between 2.6 and 4.6.
   }
The constancy of the near-IR extinction law implies that the size
distribution of the large grains responsible for the near-IR
extinction are almost the same in all directions.
A ``universal'' power law also well describes
the near-IR polarization $P_\lambda$ which,
like the near-IR extinction, probes the large grain population,
$P_\lambda/P_{\rm max}\propto \lambda^{-1.8}$, where $P_{\rm max}$
is the peak polarization (Martin \& Whittet 1990).

However, much steeper power-laws have also been derived
for the near-IR extinction.
Stead \& Hoare (2009) determined $\alpha\approx 2.14^{+0.04}_{-0.05}$
for the slope of the near-IR extinction power law
for eight regions of the Galaxy
between $l \sim 27^{\rm o}$ and $\sim 100^{\rm o}$.\footnote{%
   Stead \& Hoare (2009) derived $\alpha$ from comparing
   the UKIDSS Galactic Plane Survey data
   with the Galactic population synthesis model data
   reddened by a series of power laws and convolved
   through the UKIDSS JHK filter profiles.
   They argued that the discrepancy (i.e., a much steeper power exponent
   of $\alpha\approx2.14$ than the typical value of
   $\alpha\approx1.6-1.8$) is due to an inappropriate
   choice of filter wavelength in conversion from color excess ratios
   to $\alpha$ and that effective rather than isophotal wavelengths
   would be more appropriate.
   Naoi et al.\ (2006) also found that the near-IR color excess ratio
   $E(\JJ-\HH)/E(\HH-\Ks)$ measured in different photometric systems
   could be appreciably different.
   }
\citet{Nishiyama09} explored the extinction law toward the Galactic center.
They derived the index of the power law $\alpha\approx1.99$.
\citet{Fritz11} found $\alpha\approx2.11$ for the Galactic center extinction.

Based on  a spectroscopic study of the 1--2.2$\mum$ extinction law
toward a set of nine ultracompact HII regions with $A_{\rm V}>15\magni$,
Moor et al.\ (2005) found some evidence that the near-IR extinction
curve may tend to flatten at higher extinction,\footnote{%
   They argued that flatter curves are most likely the result of
   increasing the upper limit of the grain-size distribution,
   although the effects of flattening as a result of unresolved
   clumpy extinction cannot be ruled out.
   }
and no evidence of extinction curves significantly {\it steeper} than
the standard law (even where water ice is present).
Naoi et al.\ (2006) determined the near-IR color excess ratio
$E(\JJ-\HH)/E(\HH-\Ks)$, one of the simplest parameters for
expressing the near-IR extinction law, for L1688, L1689,
and L1712 in  the $\rho$ Oph cloud, and Cha I, Cha II, and Cha III
in the Chamaeleon cloud. They found that $E(\JJ-\HH)/E(\HH-\Ks)$
changes with increasing optical depth, consistent with grain growth
toward the inside of the dark clouds.

\subsection{Mid-IR Extinction: Universally Flat?}
The mid-IR extinction at $3\mum <\lambda<8\mum$
(in between the power-law regime at $\simali$1--3$\mum$
and the 9.7$\mum$ silicate Si--O stretching absorption feature)
is not well understood. The determination of the mid-IR extinction law
has been more difficult because this wavelength range
is best accessed from space.

Rieke \& Lebofsky (1985) argued that the near-IR power law of
$A_\lambda\propto \lambda^{-1.61}$ extends out to $\simali$7$\mum$
before the extinction increases again
toward the $9.7\mum$ silicate absorption peak.
Bertoldi et al.\ (1999) and Rosenthal et al.\ (2000)
studied the rotation-vibrational emission lines of H$_2$
of the Orion Molecular Cloud (OMC) obtained by ISO/SWS.
They found that the relative line intensities of H$_2$ appear
consistent with a power-law IR extinction of
$A_\lambda\propto \lambda^{-1.7}$
extending from $\simali$2$\mum$ to $\simali$7$\mum$
(with an additional water ice absorption feature peaking at 3.05$\mum$).

However, based on the atomic H recombination lines
toward Sgr A$^{\ast}$ in the Galactic center obtained by ISO/SWS,
Lutz et al.\ (1996) derived the extinction toward the Galactic center
between 2.5$\mum$ and 9$\mum$ by comparing the observed and
expected intensities of these lines. They found that the Galactic
center extinction shows a flattening of $A_\lambda$ in the wavelength
region of $3\mum <\lambda < 9\mum$, clearly lacking the pronounced
minimum in the $\simali$4--8$\mum$ region expected for the standard
silicate-graphite model (see Draine 1989).\footnote{%
   The silicate-graphite model for the diffuse cloud with
   $R_{\rm V}=3.1$ predicts the mid-IR extinction to be
   a simple continuation of the near-IR power law
   of $A_\lambda \propto \lambda^{-1.71}$
   (Weingartner \& Draine 2001; hereafter WD01).
   The extinction law of Lutz et al.\ (1996) for the Galactic center
   is in better agreement with the same model
   but for $R_{\rm V}\approx 5.5$ which is more suitable
   for dense regions (see Draine 2003).
   }
This was later confirmed by Lutz (1999), Nishiyama et al.\ (2009),
and Fritz et al.\ (2011).

Based on data from the {\it 2MASS} survey
and the {\it Spitzer}/Galactic Legacy Infrared Mid-Plane Survey
Extraordinaire ({\it GLIMPSE}) Legacy program,
Indebetouw et al.\ (2005) determined the $\simali$1.25--8$\mum$
extinction laws photometrically from the color excesses of background
stars for two very different lines of sight in the Galactic plane:
$l=42\arcdeg$ which points toward a relatively quiescent region,
and $l=284\arcdeg$ which crosses the Carina Arm and contains RCW~49,
a massive star-forming region.
The extinction laws $A_\lambda/A_\Ks$ derived
for these two distinct Galactic plane fields are remarkably similar:
both show a flattening across the 3--8$\mum$ wavelength range
and lie above  the curve extrapolated from $\lambda<3\mum$.
They are consistent with that derived by Lutz et al.\ (1996)
for the Galactic center, in spite of the large differences
in method and sightlines.
Even the low-density lines of sight in the Galactic midplane
follow the same flat trend (see Zasowski et al.\ 2009).

Flaherty et al.\ (2007) obtained the mid-IR
extinction laws in the {\it Spitzer}/IRAC bands
for five nearby star-forming regions:
the Orion A cloud, NGC 2068/71, NGC 2024/23,
Serpens and Ophiuchus. The derived extinction laws
at $\simali$4--8$\mum$ are flat, even flatter than
that of Indebetouw et al.\ (2005).\footnote{%
   They interpreted this as that the extinction laws
    they derived are for dense molecular clouds
    while that of Indebetouw et al.\ (2005) are for diffuse clouds.
    }
We note that almost all recent studies based on {\it Spitzer}/IRAC data
show that the mid-IR extinction law for most regions
departs from the power-law of $A_\lambda \propto \lambda^{-\alpha}$
(with $\alpha\sim$\,1.6--1.8) around 3$\mu$m and becomes flat
until the silicate feature around 9.7$\mum$.

All these observations appear to suggest
an ``universal'' extinction law in the mid-IR,
with little dependence on environments.
Chapman et al.\ (2009) computed the extinction law
at 3.6--24$\mum$ for three molecular clouds:
Ophiuchus, Perseus, and Serpens.
However,
they found that the shape of the mid-IR extinction law
in all three clouds
appears to vary with the total dust extinction:
for $A_\Ks < 0.5\magni$, the extinction law is well-fit
by the WD01 $R_{\rm V} = 3.1$ diffuse cloud dust model;
as the extinction increases, it gradually flattens;
for $A_\Ks > 1\magni$, it becomes more consistent
with the WD01 $R_{\rm V} = 5.5$ model
and that of Lutz et al.\ (1996) for the Galactic center.

Similar conclusions were drawn by McClure (2009)
who used \emph{Spitzer}/IRS observations of G0--M4\,III stars
behind dark clouds to construct empirical extinction laws
at 5--20$\mum$ for $0.3 < A_\Ks<7\magni$:
for $A_\Ks<1\magni$ the extinction law appears similar to
the $R_{\rm V}=3.1$ diffuse cloud extinction curve;
for $A_\Ks>1\magni$, it lies closer to the WD01 $R_{\rm V}=5.5$ curve.
McClure (2009) attributed the change of the shape of the IR
extinction law (as well as the 9.7$\mum$ silicate absorption
profile) to grain growth via coagulation and the presence of ice
in dense molecular clouds (at  $A_\Ks>0.5\magni$
or $A_{\rm V}> 3\magni$; Whittet et al.\ 2001).

\citet{Cambresy11} explored the variations of
the mid-IR extinction law within the Trifid nebula
by measuring the extinction in
the 3.6, 4.5, and 5.8$\mum$ {\it Spitzer}/IRAC bands.
They found a flattening of the extinction law at a
threshold of $A_{\rm V}\approx 20\magni$:
below this threshold the extinction law is as expected
from models for $R_{\rm V}=5.5$ whereas above
$\simali$20$\magni$ of visual extinction, it is flatter.

Based on the data obtained from the \textit{Spitzer}/GLIPMSE Legacy
Program and the {\it 2MASS} project, Gao, Jiang, \& Li (2009; hereafter GJL09)
derived the extinction in the four IRAC bands (relative to the
{\it 2MASS} $\Ks$ band) for 131 GLIPMSE fields along the
Galactic plane within $|l|\leq65^{\rm o}$, using red giants and red
clump giants as tracers.
As a whole, the mean extinction in the IRAC bands
(normalized to the 2MASS $\Ks$ band)  exhibits little variation
with wavelength (i.e., somewhat flat or gray), consistent with
the WD01 $R_V=5.5$ model extinction.
As far as individual sightline is concerned, however,
the wavelength dependence of the mid-IR interstellar
extinction $A_{\lambda}/A_\Ks$ varies from one sightline
to another.
GLJ09 also demonstrated the existence of systematic variations
of extinction with Galactic longitude which appears to
correlate with the locations of spiral arms:
the dips of the extinction ratios $A_\lambda/A_\Ks$
coincide with the locations of the spiral arms.
GJL09 explained this in terms of the concentration of interstellar gas
and dust at the inner edges of spiral arms which causes an increase of
grain size and dust number density.

Using data from \emph{2MASS} and \emph{Spitzer}/IRAC
for G and K spectral type red clump giants,
Zasowski et al.\ (2009) also found global longitudinal variations
of the 1.2--8$\mum$ IR extinction over $\simali$150$^{\rm o}$
of contiguous Galactic mid-plane longitude;
more specifically, they found strong, monotonic
variations in the extinction law shape as a function
of angle from the Galactic center, symmetric on either
side of it: the IR extinction law becomes increasingly
steep as the Galactocentric angle increases,
with identical behavior between $l$\,$<$\,180$^{\rm o}$
and  $l$\,$>$\,180$^{\rm o}$.

However, other studies found no evidence for a flattening
of the extinction law as a function of column density.
\citet{Roman-Zuniga07} obtained the extinction law
$A_\lambda/A_\Ks$ between 1.25 and 7.76$\mum$
to an unprecedented depth
in Barnard 59, a star-forming dense core located in the Pipe nebula,
by combining sensitive near-IR data obtained
with ground-based imagers on ESO NTT/VLT
with space mid-IR data acquired with Spitzer/IRAC.
However, they found no significant variation
of the extinction law with depth.\footnote{%
   The mid-IR extinction law is shallow and
    agrees closely with the WD01 $R_{\rm V}=5.5$ dust extinction
    model with a grain size distribution favoring larger grains than
    those diffuse clouds,  possibly due to the effect of
    grain growth in dense regions.
    }
This has been confirmed by Ascenso et al.\ (2013) who derived
the mid-IR extinction law toward
the dense cores B59 and FeSt 1-457 in the Pipe nebula
over a range of depths between $A_{\rm V}\sim10\magni$
and $A_{\rm V}\sim50\magni$,
also using a combination of Spitzer/IRAC, and ESO NTT/VLT data.
They found no evidence for a dependence of the extinction law
with depth up to $A_{\rm V}\sim50\magni$ in these cores.
The mid-IR extinction law in both cores departs significantly
from a power-law between 3.6 and 8$\mum$, suggesting that these
cores contain dust with a considerable fraction of large dust grains.

\subsection{This Work}
As discussed in previous sections, it is now clear that the
UV/optical/IR extinction varies from one sightline to another.
This variation is mainly caused by the change of
the grain size distribution in different environments.
While the UV/optical extinction is relatively well understood,
our understanding of the near- and mid-IR extinction is still
somewhat poor and controversial, despite that in this spectral
domain many advances have been made in the past few years.
A better understanding of the regional variation of
the near- and mid-IR extinction law will allow a more accurate
reddening correction of the photometric and spectroscopic measurements.
This is also crucial for a complete description of
the varying dust properties across the Milky Way.

As summarized in \S1.3, the IR extinction law is generally estimated
in molecular clouds as a whole and the discrepancies
between the different results are interpreted as variations
of the dust properties from cloud-to-cloud due to the environment.
However, most sightlines probably consist of a mixture of different
types of clouds (e.g., a concentration of discrete clouds),
instead of isolated, homogenous clouds.
Within a cloud, the gas density is expected to increase from
the envelope to the core and it is likely that a dense cloud has
an ``onion-like'' structure, with dense cloud material in the center,
surrounded by translucent gas, which is in turn surrounded by more
diffuse gas.
Therefore, the variation of the dust properties
(and hence the extinction) within a cloud is expected.
Dense clouds sample different physical
conditions in the ISM and
can be probed on large scales and for various
astrophysical environments.

With the wealth of available deep data
in the near- and mid-IR it is now
possible to investigate the variation of
the dust extinction within individual clouds.
To reveal whether and how the mid-IR extinction law
relates to the interstellar environment,
in this work we explore the possible variations
of the mid-IR extinction within the Coalsack nebula,
a nearby starless dark cloud.
We select individual regions across the nebula complex
that span a wide variety of environments,
from dense clouds to diffuse regions (see \S2).
Our analysis relies on the {\it 2MASS} and {\it Spitzer}/GLIMPSE surveys.
Our method to derive the extinction law from 3.6 to 8.0$\mum$
is based on the ``color-excess'' method with red giants as
the extinction tracer (see \S3).
We report in \S4 the derived $A_\lambda/A_\Ks$, the extinction
in the IRAC bands (relative to the {\it 2MASS} $\Ks$ band)
for the selected regions, as well as the mean mid-IR extinction.
In \S5 we discuss the regional variations of the derived IR extinction.
We summarize our major conclusions in \S6.

\section{The Coalsack Nebula}
The Coalsack nebula is the most prominent,
isolated dark cloud in the southern Milky Way
(Tapia 1973, Bok 1977).
Located around the Galactic longitude of $l\approx 303^{\circ}$
in the Galactic plane, it subtends an angle of $\simali$6$^{\rm o}$
in diameter on the sky. The distance to the Coalsack nebula is
estimated to be $\simali$150\,pc from photometric studies,
which means a linear extent of $\simali$15\,pc
(see Cambr\'esy et al.\ 1999).
It is characterized by complex, filamentary molecular structures
which contain many dark cores
and its total mass is estimated to be
$\simali$3500\,M$_{\odot}$ (see Nyman 2008).
A cloud of this size and mass would be expected to contain young
stars, but unlike other typical star-forming clouds such as Taurus
and $\rho$ Ophiuchus, the Coalsack nebula appears completely
devoid of star formation activity:
it lacks the usual signposts of star formation activity,
such as emission-line stars, Herbig-Haro objects,
and embedded IR sources
(Nyman et al.\ 1989, Bourke et al.\ 1995,
Kato et al.\ 1999, Kainulainen et al.\ 2009).
This suggests that the Coalsack nebula
may be a young molecular cloud complex
in the earliest stages of evolution.
The globules within the Coalsack,
including Tapia's Globules 1, 2, and 3,
the darkest and densest cores,
are apparently all starless
(and not centrally condensed)
and may also be in the early phases of development.
Moreover, the fraction of dense gas in the Coalsack complex is
considerably smaller than that of typical star forming cloud
complexes, as indicated by a $^{13}$CO
emission line survey (Kato et al.\ 1999),
also suggesting that the Coalsack nebula
may not be forming stars.

The visual extinction $\AV$ over the cloud varies
between $\simali$1 and $\simali$3\,mag
(see Nyman 2008).
\citet{Gegorio-Hetem88} derived the average visual extinction
of the cloud to be $\AV\approx5\magni$
through the method of star counts.
However, $\AV$ differs internally and can be much higher
in small condensations and globules,
reaching above 20$\magni$ in some regions
(Tapia 1973, Bok 1977).
The highest visual extinction in the cloud centering around
($l=301^{\circ}, b=-1^{\circ}$) contains Tapia's Globule 2,
a potential star-forming region.
With a large size and the complexity,
the Coalsack nebula presents multiple environments,
from opaque dense cores, translucent regions,
to more extended diffuse components.
Therefore, the Coalsack nebula is an ideal target
for probing the regional variation of interstellar extinction.

\begin{figure}
\centering
 \includegraphics[angle=0,width=6.5in]{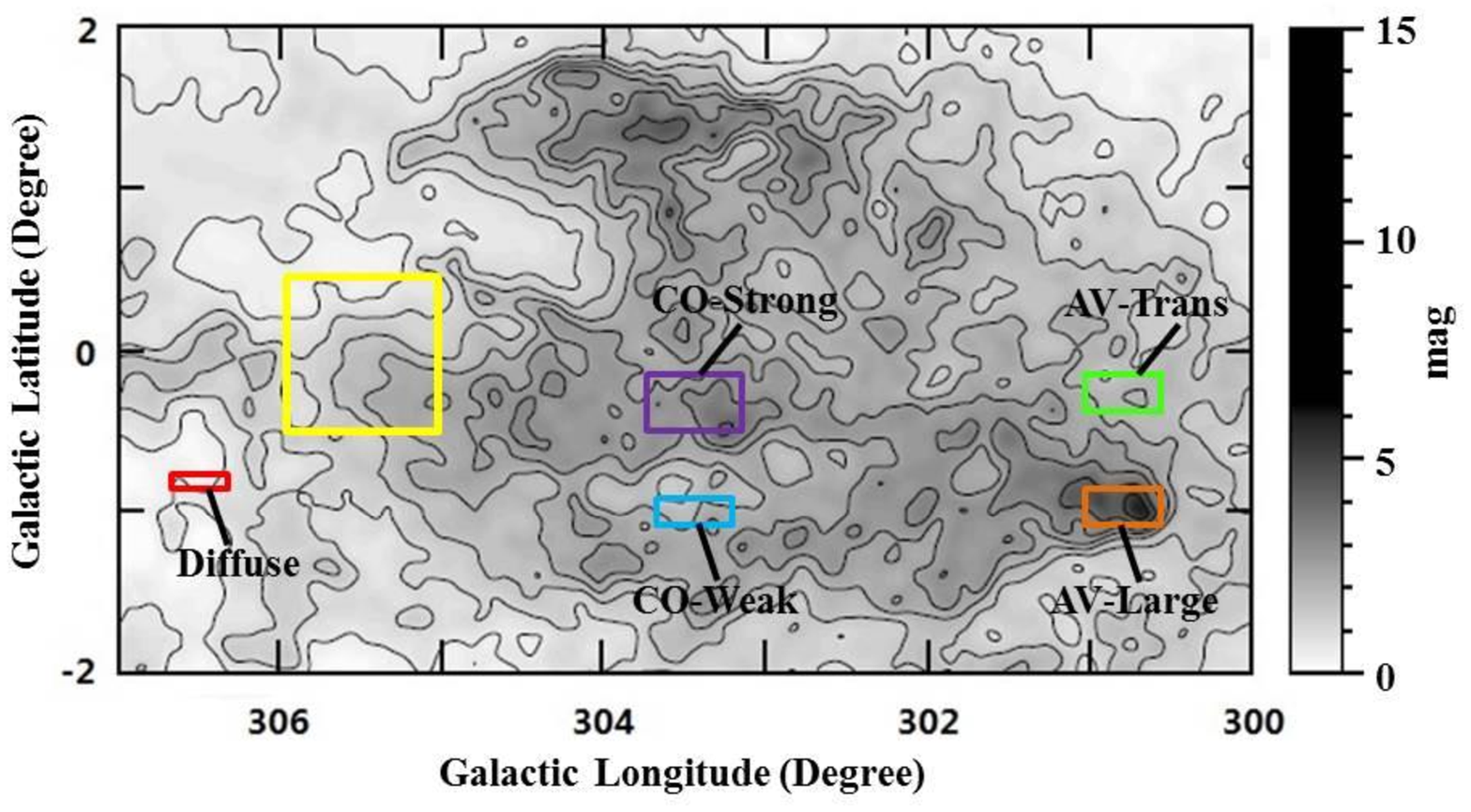}
\caption{\footnotesize
               \label{fig:extmap}
               The visual extinction map of the Coalsack nebula
               of Dobashi et al.\ (2005). The resolution of the map is
               $\simali$6$^{\prime}$.
              Orange box: the ``AV--Large'' region which, with $\AV >
              10\magni$, has the highest visual extinction in the map.
              Green box: the ``AV--Trans'' region for which the visual
              extinction is much lower (with $\AV<3\magni$).
              Also marked are the other selected regions
             (see Figures~\ref{fig:iremmap},\,\ref{fig:comap}).
              }
\end{figure}

\begin{figure}
\centering
 \includegraphics[angle=0,width=6.5in]{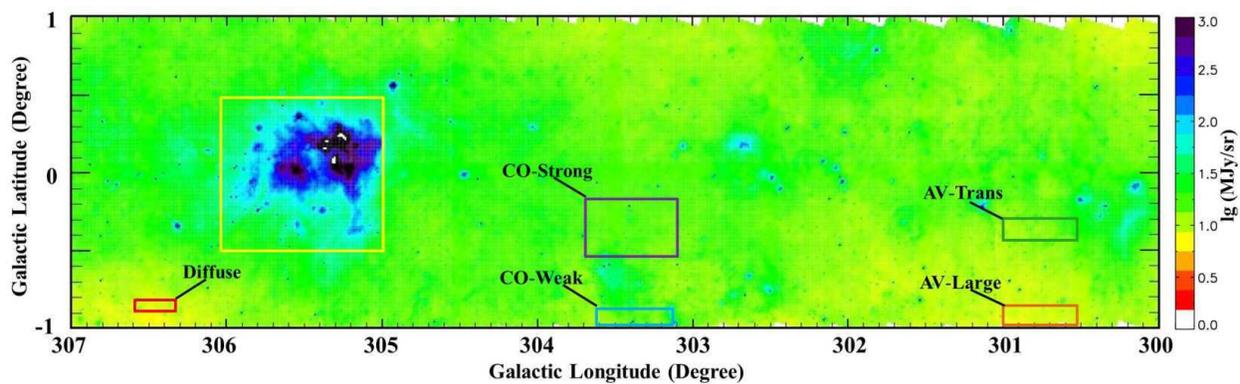}
\caption{\footnotesize
               \label{fig:iremmap}
               The \textit{Spitzer}/MIPS 24$\mum$ image of the
               Coalsack nebula. The resolution of the mosaic image
               is $\simali$6$^{\prime}$ per pixel.
               The blue and violet filaments in the yellow box represent the
               brightest region in the 24$\mum$ emission map
               (however, it is not a part of the Coalsack nebula but a
               star-forming background cloud).
               }
\end{figure}

\begin{figure}
\centering
 \includegraphics[angle=0,width=6.5in]{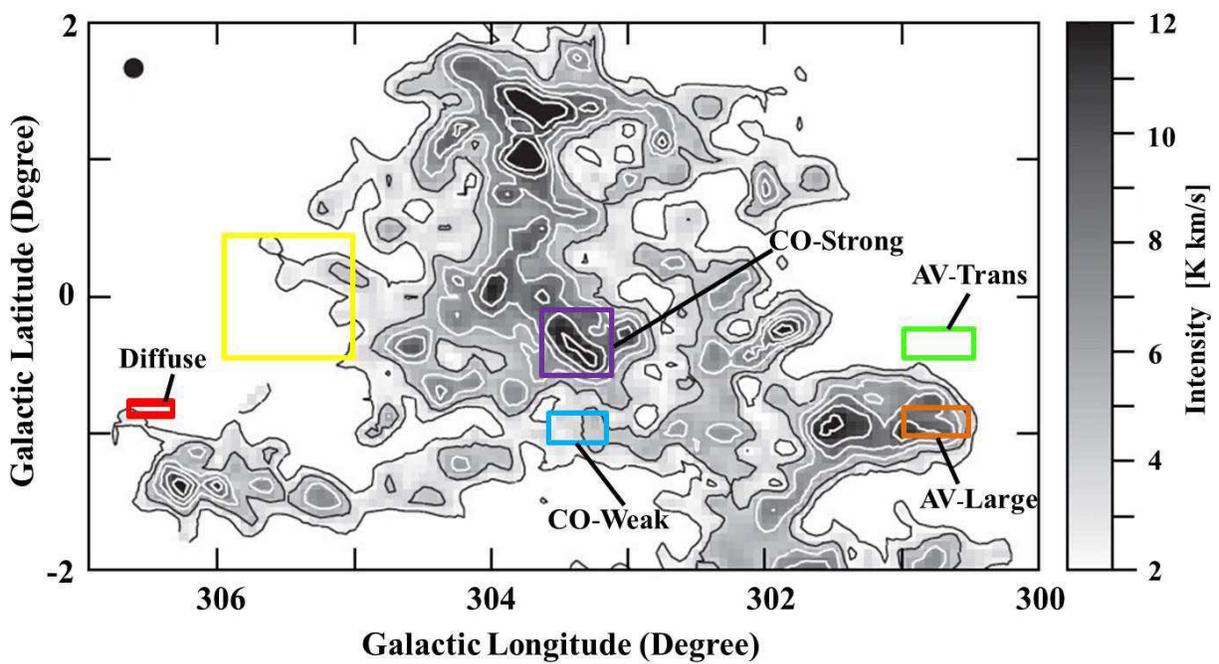}
\caption{\footnotesize
               \label{fig:comap}
               Map of the velocity-integrated CO\,(1-0) emission
               intensity of Coalsack \citep{Nyman89}.
               Contours are from 2$\K\km\s^{-1}$ to 12$\K\km\s^{-1}$
               in steps of 1$\K\km\s^{-1}$.
               Violet box: the ``CO--Strong'' region.
               Blue box: the ``CO--Weak'' region.
               The other selected regions are also marked.
               }
\end{figure}

To characterize the ISM environment,
three maps of the Coalsack nebula
are utilized independently:
(i) the visual extinction map,
(ii) the \textit{Spitzer}/MIPS 24$\mum$ emission map, and
(iii) the CO gas emission contour.
Figures~\ref{fig:extmap}--\ref{fig:comap}
display these maps of the Coalsack nebula
for the selection of regions
representative of different environments.
\begin{enumerate}
\item {\it Visual Extinction} ($\AV$). The extinction
          depends on the dust column density $N_{\rm d}$
          and the optical properties of the dust.
          Let $a$ be the mean grain size
          and $C_{\rm ext}(a,\lambda)$ be the extinction cross section
          of the dust of size $a$ at wavelength $\lambda$.
          The extinction is
          $A_\lambda = 1.086\,N_{\rm d}\,C_{\rm ext}(a,\lambda)$.
          Apparently,  a large $A_\V$ often implies a dense cloud
          (although it is also possible that it is just a pile-up of
          many diffuse clouds along the line of sight).
          \citet{Dobashi05} presented a comprehensive
          catalog of dark clouds, in which the Coalsack nebula is
          labelled as No.\,1867.
          Figure~\ref{fig:extmap} is the $\AV$ map around the Coalsack region
          with an angular resolution of $\simali$6$^{\prime}$
          for which the lowest contour is 0.5\,mag,
          with an interval of 0.5\,mag
          in the range of $\AV<3.0\magni$
          and of 1.0\,mag in the range of $\AV>3.0\magni$.
          As described in \S1.1, we distinguish different regions in
          terms of $\AV$: (1) the region with the largest $\AV$
          ($>$\,10\,mag), designated as ``AV--Large''
          in Tables~\ref{tab:5regions} and \ref{tab:colorexcess}
          and marked as an orange square in Figure~\ref{fig:extmap},
          is chosen as a representative of dense clouds;
          (2) the region with moderate $\AV$ ($<$\,3\,mag),
          designated as ``AV--Trans''
          in Tables~\ref{tab:5regions} and \ref{tab:colorexcess}
          and marked as a green square in Figure~\ref{fig:extmap},
          is chosen as a representative of translucent clouds.
\item {\it Dust IR Emission}. Let $T$ be the dust temperature
          (or the mean temperature averaged over a distribution of
           temperatures for stochastically heated dust).
           The dust emission intensity is
           $I_\lambda \propto N_{\rm d}\,C_{\rm abs}(a,\lambda)\,B_\lambda(T)$
           where $B_\lambda(T)$ is the Planck function of temperature
           $T$ at wavelength $\lambda$.
           As the Coalsack nebula is starless, the dust is externally
           illuminated by the general interstellar radiation field.
           The dust in dense cores is colder than that in translucent
           or diffuse regions because of the attenuation of the
           external UV radiation. Therefore, regions bright in the IR
           must be dense. We adopt the \emph{Spitzer}/MIPS 24$\mum$
           image of the Coalsack nebula due to its high sensitivity
           and high spatial resolution.\footnote{%
              Downloaded from the \emph{Spitzer} Data Archive
               {\sf
                http://irsa.ipac.caltech.edu/data/SPITZER/MIPSGAL/images/
               }.
              }
           The brightest region marked as a yellow square
           in Figure~\ref{fig:iremmap}, locating at
           $305.0^{\circ}<l<306.0^{\circ}$ and
           $-0.5^{\circ}<b<0.5^{\circ}$,
           is supposed to be very dense.
           However, it is not associated with the Coalsack nebula.
           It is a star-forming background cloud (see \citealt{CP04}).
           Neither the ``AV--Large'' region nor the ``AV--Trans''
           region exhibits appreciable 24$\mum$ emission.
\item {\it CO Emission}. In comparison with $\AV$
          and the 24$\mum$ dust emission, CO as a gas molecule
          is a relatively indirect tracer of dust.
          If the gas is well mixed with the dust
          as indicated by the tight correlation
          between the HI intensity and $\AV$,
          the intensity of the CO emission lines
          should be proportional to the dust density
          (although this is complicated by the fact that
          the intensity of the CO emission line also depends on
          the excitation temperature).
          Indeed, Zasowski et al.\ (2009) used the $\rm {}^{13}$CO
          (J\,=\,1$\to$0)  line to trace dense interstellar clouds
          (i.e., only dense gas can show up in the $^{13}$CO line
          emission).
          Figure~\ref{fig:comap} presents the integrated CO\,(1-0)
          intensity map of the Coalsack nebula with a resolution
          of $\simali$8$^{\prime}$ \citep{Nyman89}.
          Complex structures are clearly seen in the CO map.
          Guided by the integrated CO line intensity map,
          we select two regions in the Coalsack nebula:
          the ``CO--Strong'' region which has the strongest
          line intensity (The CO emission line intensity can reach 6\,K\,km/s),
          and the ``CO--Weak'' region which is weak in CO emission
          (The intensity of the CO emission line is less than 2\,K\,km/s).
          These two regions should correspond to dense and translucent
          environments, respectively.
          They are marked as violet and blue squares
          in Figure~\ref{fig:comap}, respectively.
\end{enumerate}

As shown in Figure~\ref{fig:comap}, the selections based on
the visual extinction $\AV$, the 24$\mum$ dust emission,
and the CO emission are generally consistent, but discrepancy
also exists.
The ``AV--Large'' region is strong in the CO line emission,
although not the strongest.
The ``AV--Trans'' region is weak in CO emission:
with an intensity of $\simali$1$\K\km\s^{-1}$
(see Figure~1 of \citealt{Nyman89})
it does not show up in Figure~\ref{fig:comap}
for which the lowest contour
is 2$\K\km\s^{-1}$.
All these regions do not seem to be very bright in the
{\it Spitzer}/MIPS 24$\mum$ map (see Figure~\ref{fig:iremmap}),
especially the ``CO--Strong'' region
or the ``AV--Large'' region.
This is because the 24$\mum$ dust emission is not only proportional to
the dust column density $N_{\rm d}$ but also non-linearly proportional
to the dust temperature $T$, while $T$ depends on $N_{\rm d}$ as well:
$T$ is lower in denser regions because of the attenuation
of the illuminating interstellar UV radiation.
Therefore, the densest regions (e.g., ``AV--Large'') are not
necessarily the brightest in the {\it Spitzer}/MIPS 24$\mum$ emission
map because the darkest and densest cores have the lowest $T$.
Because of this, the visual extinction $\AV$ appears to be the best
to characterize dense, translucent, and diffuse regions.

Finally, we select a diffuse region
(designated ``Diffuse''; see Figures~\ref{fig:extmap}--\ref{fig:comap})
near the east edge of the Coalsack nebula
where there is essentially no dust extinction
($\AV<1\magni$; see Figure~\ref{fig:extmap}),
very little CO emission
(the intensity of CO emission is less than 1\,K\,km/s),
and undetectable 24$\mum$ emission (see Figure~\ref{fig:comap}).

To summarize, in total five regions are selected
(see Table~\ref{tab:5regions}):
two dense regions
(``AV--Large'', and ``CO--Strong''),
two translucent regions
(``AV--Trans'', and ``CO-Weak'') ,
and one diffuse region (``Diffuse'')
within the Coalsack nebula.
These regions cover various interstellar environments,
from diffuse, translucent to dense.

We also note that the Coalsack nebula is located
at a Galactic latitude close to $\simali0^{\rm o}$,
for a selected sightline the extinction could be
contaminated by background clouds.
For example, the ``AV--Trans'' region,
with $\AV\simlt3\magni$
and $A_{\rm J}/A_{\rm K_{\rm s}}\approx 3.28$
(see \S\ref{sec:results} and Table~\ref{tab:irext}),
is expected to be reddened by an amount of
$E(\JJ-\Ks)\approx0.8\magni$
(assuming $A_{\rm V}/A_{\rm K_{\rm s}}\approx 8.5$),
while Figures~\ref{fig:cmdiagram},\ref{fig:ccdiagram}
show that the $\JJ-\Ks$ colors of some red giants
in the ``AV--Trans'' region exceed $\simali$3$\magni$,
implying that background clouds could contribute $\simgt$1$\magni$
to the reddening of these sources (assuming an intrinsic color
of $\JJ-\Ks\approx 1.2\magni$ for red giants).
To quantitatively estimate the contributions of background clouds
to the extinction, we calculate the $E(\JJ-\Ks)$ values
from $\AV$ and compare them with that from
the $\JJ$ vs. $\JJ-\Ks$ color-magnitude diagrams
(e.g., see Figure~\ref{fig:cmdiagram}).
It is found that the extinction of
the ``Diffuse'', ``AV--Large'' and ``CO--Weak'' regions
is mainly from these regions, with little contribution
from background clouds,
while the ``CO--Strong'' region is
similar to the ``AV--Trans'' region,
with background clouds contributing
an additional reddening
of $E(\JJ-\Ks)\simgt2\magni$.

\begin{figure}
\centering
 \includegraphics[angle=0,width=6.5in]{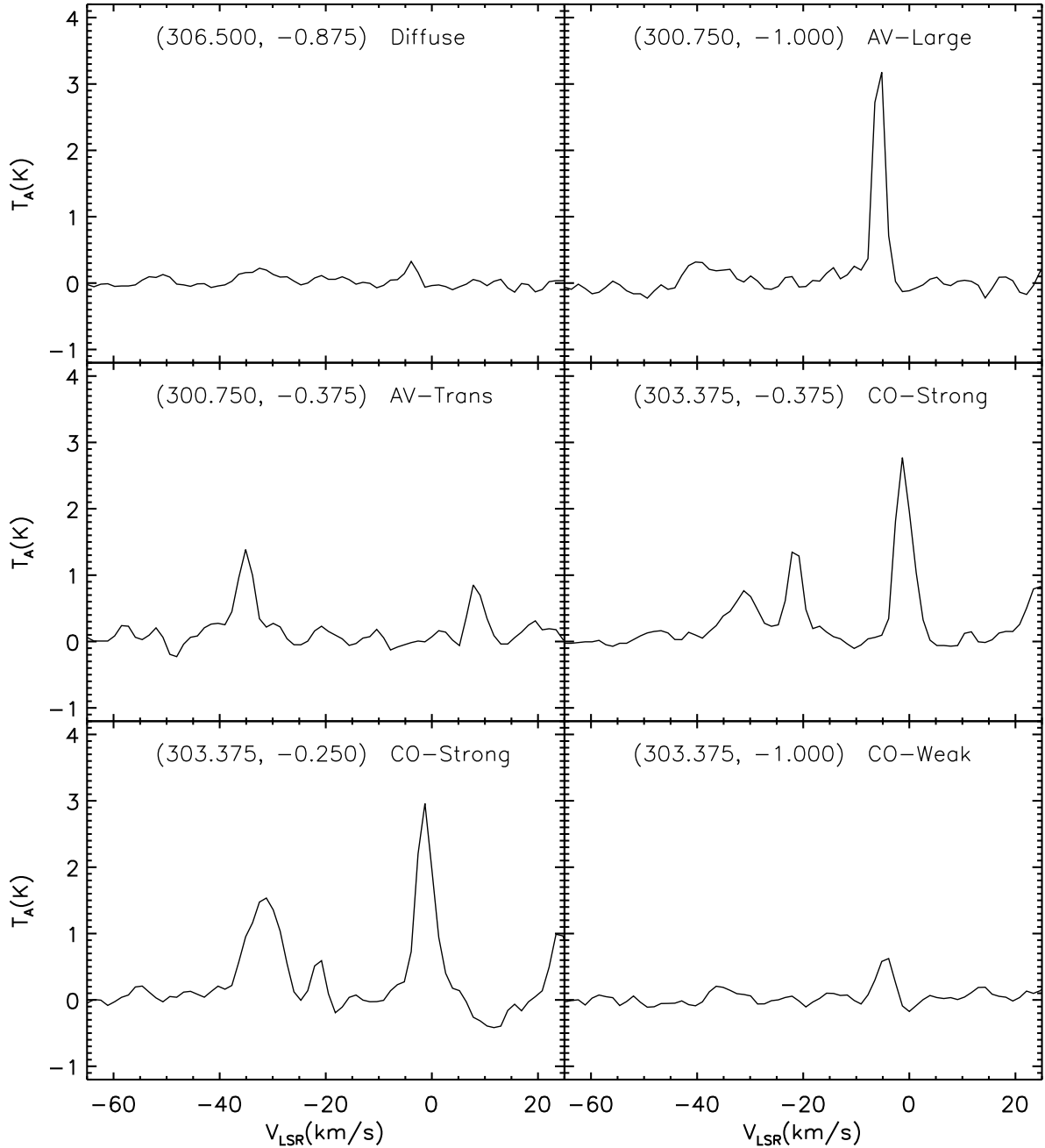}
\caption{\footnotesize
         \label{fig:CO.Line}
         Antenna temperature $T_{\rm A}$
         vs. the CO velocity $V_{\rm LSR}$
         along the lines of sight toward
         the five selected regions in the Coalsack nebula.
         The name and coordinate of each region are labelled.
         For the ``CO--Strong'' region, two diagrams are shown,
         corresponding to two close regions
         which have the same galactic longitude
         but different galactic latitude.
         The CO data are taken from Nyman et al.\ (1989).
         }
\end{figure}

\begin{figure}
\centering
 \includegraphics[angle=0,width=6.5in]{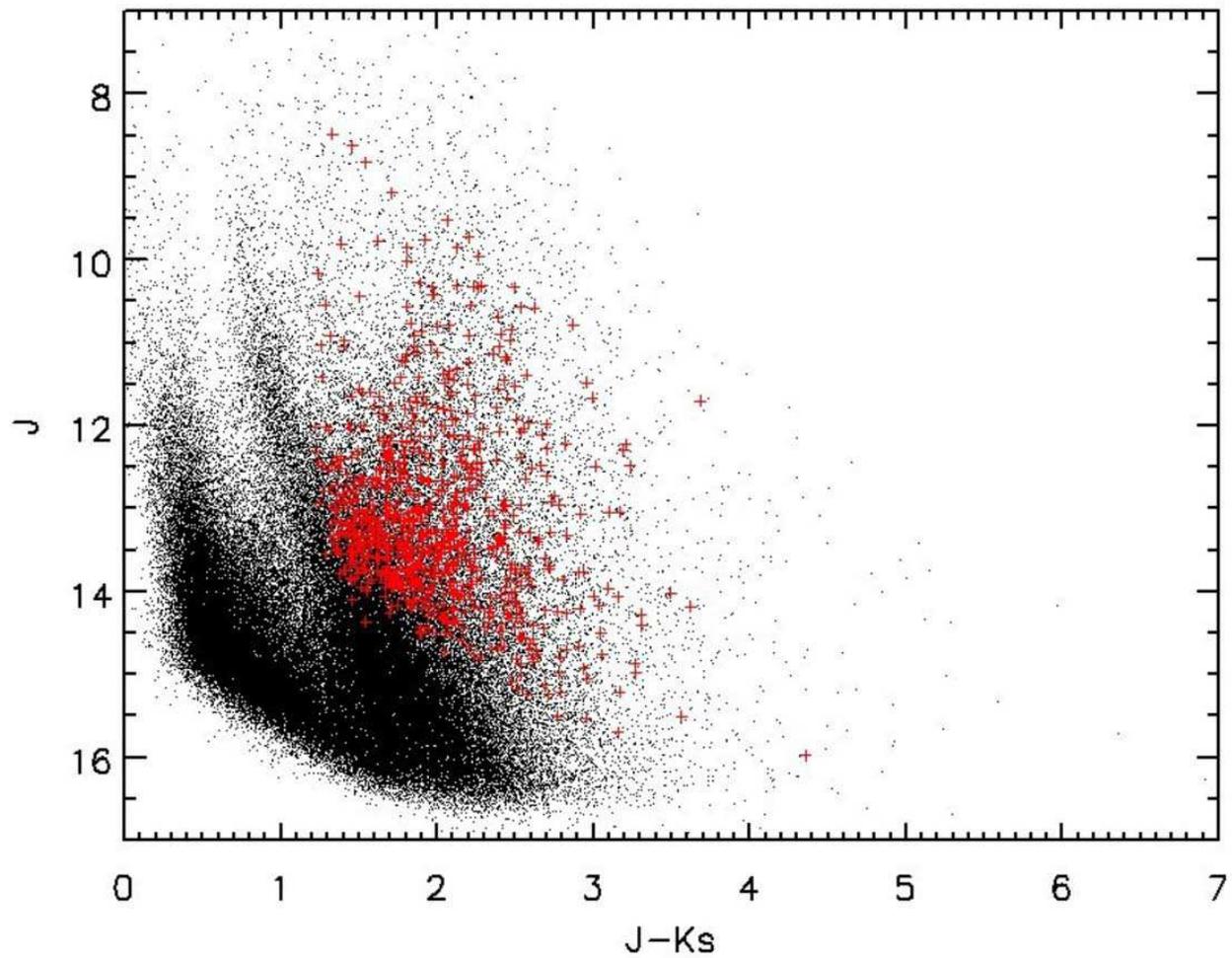}
\caption{\footnotesize
              \label{fig:cmdiagram}
             The $\left(\J-\Ks\right)$ vs. $\J$ color-magnitude diagram
             for the ``AV--Trans'' region (see Figure~\ref{fig:extmap}
             and Table~1).
             Black dots: all the sources in the GLMIC $l$300 field of
             GLIMPSE\,I which covers the ``AV--Trans'' region.
             Red crosses: the selected red giants of this region.
             }
\end{figure}

\begin{figure}
\centering
 \includegraphics[angle=0,width=6.5in]{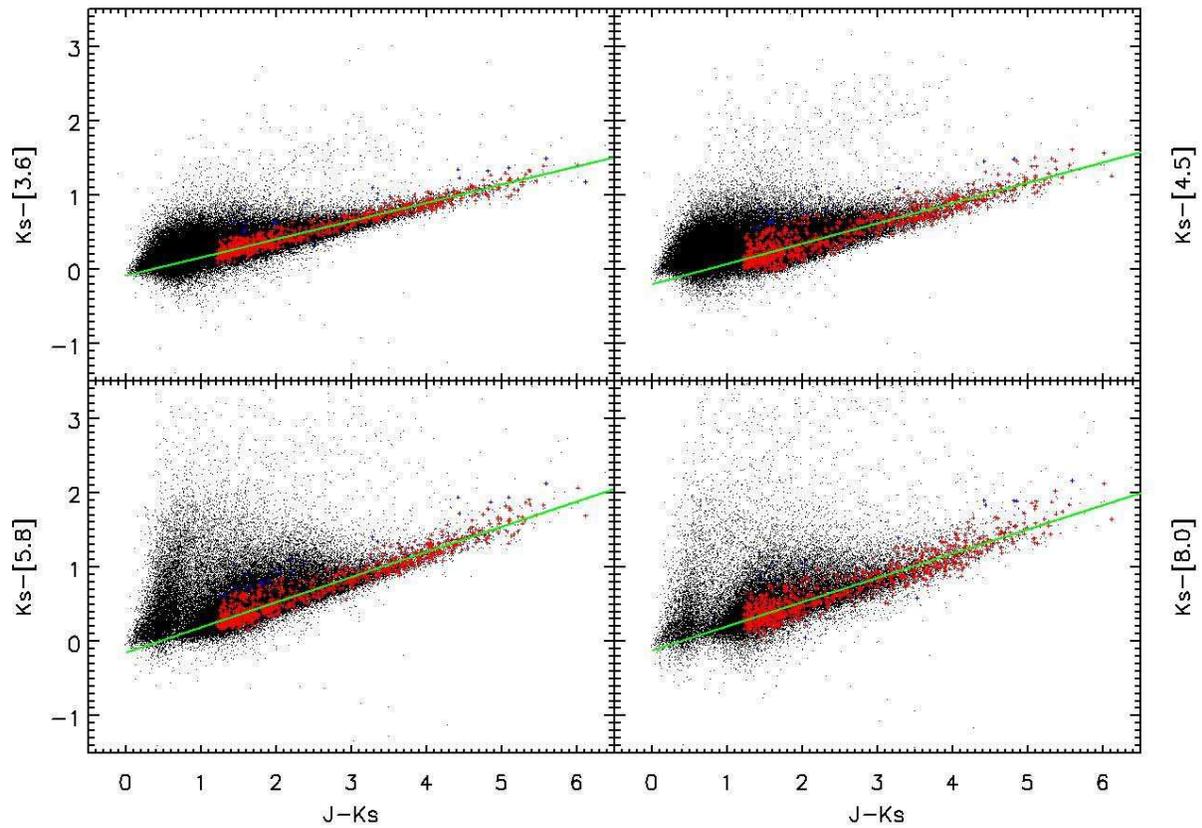}
\caption{\footnotesize
               \label{fig:ccdiagram}
              The {\it 2MASS} and {\it Spitzer}/IRAC color-color
              diagrams for the sources shown in
              Figure~\ref{fig:cmdiagram}.
              Green lines: the best fits to the data.
              Blue points: the sources which are dropped based on
              a $3\sigma$ criterion.
              }
\end{figure}

One can also gain insight into the possible background contamination
by examining the CO components of the Coalsack nebula.
In Figure~\ref{fig:CO.Line} we show the CO velocity $V_{\rm LSR}$
vs. the CO line brightness
(expressed as $T_{\rm A}$, the antenna temperature)
for these five regions.\footnote{%
   The data are taken from Nyman et al.\ (1989)
   who made a complete survey of CO J=1--0 emission
   of the Coalsack nebula with an 8 arcmin resolution
   (see {\sl http://hdl.handle.net/10904/10051V3}).
   }
It is clearly seen that the velocities
at the ``Diffuse'', ``AV--Large'' and ``CO--Weak'' regions
have only one component at $\simali-5\km\s^{-1}$
which is from the Coalsack nebula.

In contrast, there are two velocity components
toward both the ``AV--Trans'' and ``CO--Strong'' regions:
one component peaks between $-10$ to 10$\km\s^{-1}$
that is supposedly from the Coalsack nebula;
the other peaks from about $-40$ to $-20\km\s^{-1}$
which is definitely behind the Coalsack nebula.\footnote{%
  Nyman (2008) found that the background clouds
  in the near and far sides of the Carina arm have
  velocities between $-35$ to $30\km\s^{-1}$.
  }
Nevertheless, for the ``AV--Trans'' sightline,
the intensity of the CO line
at $\simali -35\km\s^{-1}$ is comparable
to that for the Coalsack nebula.
This suggests that the background cloud
may be in an environment similar to that of
the ``AV--Trans'' region and therefore, it would
have little effect on the extinction law
of the ``AV--Trans'' region.
As for the ``CO--Strong'' region, the peak intensity of
the additional component is slightly weaker than that of
the Coalsack nebula component, while the integrated
intensity is comparable.
The starlight intensity of the Coalsack nebula is likely
weaker than that of the background cloud as the Coalsack
is illuminated by the external interstellar radiation field,
while the background is at least also excited by the general
interstellar radiation field or even by an embeded or
nearby star. Therefore, the CO--Strong region of
the Coalsack nebula could be relatively denser than
the background cloud and the dust properties derived
for the CO--Strong region in this work could be contaminated
by dust in an environment which is less dense.
However, due to the lack of detailed knowledge of the radiation field
of the background cloud, it is difficult to quantitatively assess
this effect. Nevertheless, it is worth noting that the region was
selected from the CO emission intensity map
of Nyman et al.\ (1989) which was obtained by
integrating the velocity from $-10$ to $8\km\s^{-1}$ for the coalsack
nebula specifically, excluding the background cloud.

\section{Methods and Tracers} \label{method}
\subsection{The ``Color-Excess" Method}
The method of ``color excess" is used to obtain
the extinction as a function of wavelength.
The details of  the method can be found in GJL09.
In brief, this method derives the extinction from
the ratio of color excesses $E(\lambda_r-\lambda)$
and $E(\lambda_c-\lambda_r)$,
where $\lambda_r$ is the reference band,
$\lambda_c$ the comparison band,
$\lambda$ is the band investigated
(i.e., the \emph{Spitzer}/IRAC bands at [3.6], [4.5],
[5.8], and [8.0]$\mu$m).
In our study, we take the $\Ks$ band to be the reference band,
and the J band to be the comparison band.
Thus, the ``color-excess'' method calculates the ratio following
\begin{equation}\label{slope}
k_{x}\equiv\frac{E(\Ks-\lambda)}{E(\J-\Ks)}
  = \frac{(\Ks-\lambda)-(\Ks-\lambda)_0}
{(\J-\Ks)-(\J-\Ks)_0}=\frac{A_\Ks-A_{\lambda}}{A_\J-A_\Ks} ~~,
\end{equation}
where $(\Ks-\lambda)_0$ and $(\J-\Ks)_0$
are the intrinsic colors of the source.
The relative extinction $A_{\lambda}/A_\Ks$ can then be derived
from given $A_\J/A_\Ks$:
\begin{equation}\label{ext}
A_{\lambda}/A_\Ks=1+ k_x(1-A_\J/A_\Ks).
\end{equation}

Our method does not calculate the color excess ratio of individual stars,
instead, it calculates the ratio of a group of stars.
In practice, the color excess ratio $k_x$ is derived from
the slope of a linear fitting of $E(\Ks-\lambda)$ vs. $E(\J-\Ks)$.
This statistical method reduces the uncertainty by increasing
the number of sources involved.
The key of this method lies in choosing
a group of sources with homogeneous intrinsic color indices.

\subsection{Data: {\it 2MASS} and \emph{Spitzer}/GLIMPSE Surveys}
The near- and mid-IR magnitudes of the objects
are needed for the ``color-excess" method.
We obtain these data from
the {\it 2MASS} PSC catalog \citep{Skrutskie06}
and the \emph{Spitzer}/GLIMPSE PSC catalog \citep{Benjamin03}.
The {\it 2MASS} PSC provides measurements in
$\J$\,(1.24$\mu$m),
$\HH$\,(1.66$\mu$m),
and $\Ks$\,(2.16$\mu$m) bands in the whole sky.
The GLIMPSE program is a mid-IR survey program of
the inner Galactic plane in four filters
([3.6], [4.5], [5.8], [8.0]) using the IRAC camera
onboard \emph{Spitzer}. It has four cycles:
GLIMPSE\,I, GLIMPSE\,II, GLIMPSE\,3D and GLIMPSE\,360 \citep{Churchwell09}.
Furthermore, the GLIMPSE PSC (version 2.0) has already
cross-identified with the {\it 2MASS} PSC (see \citealt{Cutri03}),
bringing in great convenience to our work.
Due to the orientation of the GLIMPSE survey,
the PSC is confined to the Galactic latitude $b$ within  $\pm1^\circ$.

\subsection{Selection of the Extinction Tracers}
Red giants (RGs) are good tracers of interstellar extinction
in the IR. They have a narrow range of effective temperatures
and thus the scatter of their intrinsic color indices is small
(e.g., the intrinsic $(\J-\Ks)$ color index of RGs is $\simali$1.2\,mag
with a scatter of only $\simali$0.1\,mag; \citealt{Bertelli94, Glass99}).
Moreover, RGs are bright in the IR (e.g., their absolute magnitude
in the $\Ks$ band can reach $\simali$$-5.0$\,mag; GJL09).
In comparison, red clump stars (RCs) that are also often used as tracers
of IR extinction have $M_\Ks\sim -1.65\magni$ \citep{Wainscoat92}.
This implies that RGs can trace a distance $\simali$5 times further
than RCs. We adopt the following criteria to select RGs:
\begin{enumerate}
\item $[3.6]-[4.5]<0.6$\,mag and $[5.8]-[8.0]<0.2$\,mag.
          These restrictions exclude sources
           with intrinsic IR excesses from circumstellar dust
           \citep{Flaherty07}
           such as pre-main-sequence stars,
           asymptotic giant branch stars (AGBs)
           and young stellar objects (YSOs; \citealt{Allen04}).\footnote{%
              However, \citet{Marengo07} argued that
              in the IRAC bands, the colors of AGB stars
              are similar to that of RGs, and therefore some AGB stars
              may contaminate the selected samples.
              Fortunately, AGB stars are much less numerous than RGs
              due to their relatively short lifetime.
              }
\item $\J-\Ks>$\,1.2\,mag and $\HH-\Ks>$\,0.3\,mag.
          \citet{Marshall06} found that the dwarf population is
           on the left side of the $\Ks$ vs. $\J-\Ks$ diagram,
           and the line near $\J-\Ks$\,=\,0.9 can be used
           to exclude dwarf stars. We follow GJL09 to adopt
           the (more restrict) criteria of $\J-\Ks>\,1.2$
           and $\HH-\Ks>\,0.3$ to exclude foreground dwarfs.
\item S/N\,$\ge$\,10.
          To guarantee the photometric quality,
          we only include the sources with S/N\,$\ge$\,10
          in all seven bands (i.e., three {\it 2MASS}
          bands and four IRAC bands).
\item $<$\,3$\sigma$ deviation from the linear fitting.
          The objects selected according to the above criteria
          are taken to fit the $\J-\Ks$ vs. $\Ks-\lambda$ relation.
          Still, there are some sources with a deviation exceeding
          3$\sigma$. These sources are rejected from
          a statistical point of view, also, most of them have large
          $\Ks-\lambda$ color excesses, suggesting the extinction
          may arise from circumstellar dust \citep{Jiang06}.
\end{enumerate}

Figure~\ref{fig:cmdiagram} shows the selected RG stars
together with other stars (e.g., dwarfs, stars with IR excess)
in the $\J-\Ks$ vs. $\J$ diagram
for the ``AV--Trans'' region.
The black dots are all the sources in the field GLMIC $l$300
of GLIMPSE\,I which covers the ``AV--Trans'' region.
The selected RGs are marked as red crosses.
The branches of the dwarf and red clump stars
are clearly seen in this diagram,  while those sources
with intrinsic IR excess are mixed with RGs,
although they have redder colors in the mid-IR bands.

\section{Results\label{sec:results}}
The calculation of $A_\lambda/A_\Ks$ from the color excess ratios
$E(\Ks-\lambda)/E(\J-\Ks)$ is straightforward (see eq.\,\ref{ext}).
Here the wavelength $\lambda$ refers to
the \emph{Spitzer}/IRAC bands
at [3.6], [4.5], [5.8], [8.0]$\mum$.
For each region listed in Table~\ref{tab:5regions},
$A_\lambda/A_\Ks$ is calculated based on the selected RGs
and the linear fitting of $\J-\Ks$ vs. $\Ks-\lambda$.
The number of RG stars that participate in
the final linear fitting is shown in
the last column of Table~\ref{tab:5regions},
all exceeding 100 sources. This guarantees that
the ``color-excess'' method
(see \S3.1) to be statistically significant.
Figure~\ref{fig:ccdiagram}, as an example, illustrates this procedure
for the ``AV--Trans'' region.
During the linear fitting (to derive $k_x$, see eq.\,1),
those sources with a deviation exceeding 3$\sigma$
are dropped and such a fitting is repeated three times.
The color excess ratios $E(\Ks-\lambda)/E(\J-\Ks)$
for all the selected regions are shown
in Table~\ref{tab:colorexcess}.

\begin{figure}
\centering
 \includegraphics[angle=0,width=6.5in]{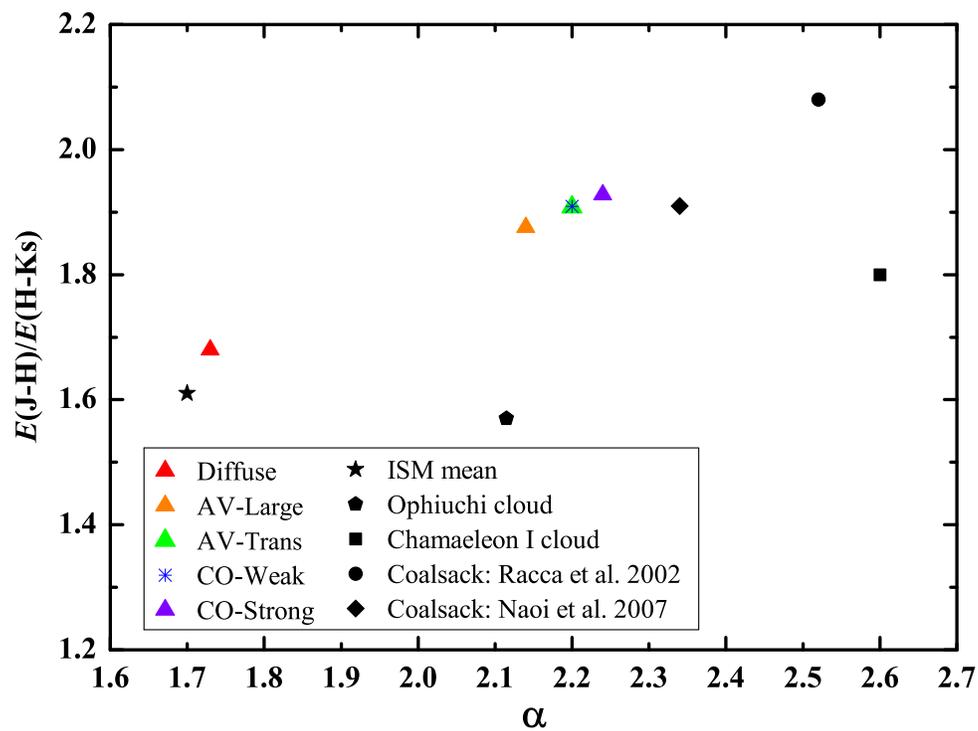}
\caption{\footnotesize
              \label{fig:ir.alpha}
              $E(\J-\HH)/E(\HH-\Ks)$ vs. $\alpha$ for all
              the five selected regions, where $\alpha$ is
              the power exponent of the near-IR extinction
              which is approximated as a power law
              $A_\lambda \propto \lambda^{-\alpha}$.
              Also shown are the $\alpha$ values estimated
              for the Ophiuchi cloud, the Chamaeleon I dark
              cloud (Kenyon et al.\ 1998, G\'omez \& Kenyon 2001)
              and the mean values averaged over diverse environments
              (Whittet 1988), as well as earlier determinations for
              the Coalsack nebula (Racca et al.\ 2002, Naoi et al.\ 2007).
              }
\end{figure}

In converting the color excess ratio
$E(\Ks-\lambda)/E(\J-\Ks)$
to the extinction ratio $A_\lambda/A_\Ks$
(see eq.\,2), the knowledge of $A_\J/A_\Ks$ for each region
is required. We therefore first derive $A_\J/A_\Ks$ for each
region based on the {\it 2MASS} photometry.
We also fit the {\it 2MASS} JHK$_{\rm s}$ extinction in terms
of a power-law $A_\lambda\sim \lambda^{-\alpha}$ and derive
the power exponent $\alpha$ for each region.
In Table~\ref{tab:irext} we tabulate $A_\J/A_\Ks$ and $\alpha$ for
all five regions.
Except for the ``Diffuse'' region, all regions have $A_\J/A_\Ks\simgt3.0$,
exceeding by $>$20\% that of \citet{RL85} for stars in the Galactic center
and of \citet{Indebetouw05} for the $l=42^{\circ}$ and $l=284^{\circ}$
sightlines in the Galactic plane:
$A_\J/A_\Ks \approx 2.52$.
In Figure~\ref{fig:ir.alpha} we plot the near-IR color excess ratio
$E(\JJ-\HH)/E(\HH-\Ks)$ over the near-IR extinction power
exponent $\alpha$ for the selected five regions as well as
for the Ophiuchi cloud, the Chamaeleon I dark cloud,
and the mean values averaged over diverse environments
(Kenyon et al.\ 1998, G\'omez \& Kenyon 2001, Whittet 1988a).

Except for the ``Diffuse'' region,
the near-IR extinction laws for all other
regions are rather steep, with $\alpha>2.0$
and reaching $\alpha\simgt2.2$ for the ``AV--Trans'',
``CO--Weak'', and``CO--Strong'' regions.
In comparison, the typical value for interstellar clouds is
$\alpha\sim 1.6-1.8$ (see \S1.3), although
$\alpha\gtsim2$ has also been reported for some regions
(e.g., see Stead \& Hoare 2009, Nishiyama et al.\ 2009; see \S1.3).
This confirms the earlier determination of
Naoi et al.\ (2007) who derived $\alpha\approx 2.34 \pm 0.01$
for  the Coalsack Globule 2, based on observations made
with the {\it Simultaneous Infrared Imager for Unbiased Survey} (SIRIUS)
on the {\it Infrared Survey Facility} (IRSF) 1.4\,m telescope at the South
African Astronomical Observatory (SAAO).

Also except for the ``Diffuse'' region,
the near-IR color excess ratios
for all other regions,
with $E(\JJ-\HH)/E(\HH-\Ks)>1.8$,
are appreciably higher than
the mean value of $1.61 \pm 0.04$
averaged over diverse environments
(Whittet 1988a).
They are also higher than
the ratios for the $\rho$ Ophiuchi cloud
and the Chamaeleon I dark cloud
($E(\JJ-\HH)/E(\HH-\Ks)$\,$\simali$1.6--1.7;
see Kenyon et al.\ 1998, G\'omez \& Kenyon 2001).
This confirms the earlier results of
Naoi et al.\ (2007) who derived
$E(\JJ-\HH)/E(\HH-\Ks)\approx1.91 \pm 0.01$
for the Coalsack Globule 2
for the extinction range $0.5 < E(\JJ-\HH) <1.8$.
Racca et al.\ (2002) also observed the Coalsack nebula
and determined $E(\JJ-\HH)/E(\HH-\Ks)\approx2.08\pm 0.03$.
Naoi et al.\ (2007) argued that steep near-IR extinction law
and the large near-IR color excess ratio might indicate little
growth of dust grains, or large abundance of dielectric
non-absorbing components such as silicates,
or both in the Coalsack Globule 2.

\begin{figure}
\centering
 \includegraphics[angle=0,width=6.5in]{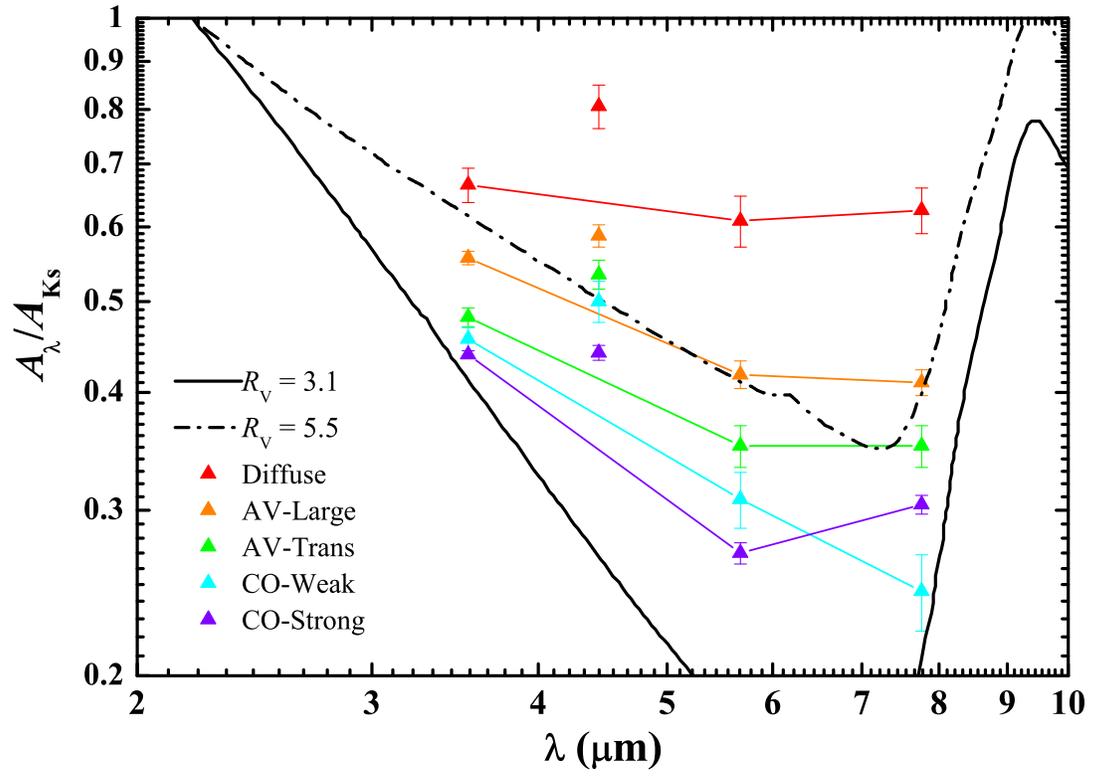}
\caption{\footnotesize
               \label{fig:irext}
               The mid-IR extinction $A_\lambda/A_\Ks$ of the five
               selected regions in the Coalsack nebula.
               For comparison, the WD01 $\Rv=3.1, 5.5$ model
               extinction curves are also shown. The error bars in this
               figure and Figure~\ref{fig:iceregions} are derived from
               the linear fitting of $E(\Ks-\lambda)/E(\J-\Ks)$.
               }
\end{figure}

With $A_\J/A_\Ks$ derived from the {\it 2MASS} dataset
for each region, we determine  the mid-IR extinction
in the four IRAC bands
(relative to that in the {\it 2MASS} $\Ks$ band)
$A_\lambda/A_\Ks$
for all the selected regions.
The results are tabulated in Table~\ref{tab:irext}
and illustrated in Figure~\ref{fig:irext}.

Most noticeably, the derived mid-IR extinction curves are
flat for all five regions (including the ``Diffuse'' region):
(i) they are all flatter than the WD01 $\RV=3.1$ model curve, and
(ii) the extinction curves of the ``Diffuse'' region and the
``AV--Large'' region are even flatter than the WD01
$\RV=5.5$ model curve.
Figure~\ref{fig:irext} also clearly shows that
the mid-IR extinction
$A_\lambda/A_\Ks$ varies with interstellar environments:
all four bands exhibit regional variations,
although to different degrees.
$A_{[3.6]}/A_\Ks$ has a minimum of $\simali$0.439
and maximum of $\simali$0.665,
with a standard deviation of $\simali$0.093,
the smallest deviation among these four bands.
$A_{[4.5]}/A_\Ks$ has a minimum of $\simali$0.441
and maximum of $\simali$0.806,
with a standard deviation of $\simali$0.140,
the largest deviation, and also the largest range of variation.
The other two bands, [5.8] and [8.0], exhibit similar variations,
also with appreciable deviations.
This confirms that the mid-IR extinction law is not universal
(see \S1.4).

The mid-IR extinction curve for the ``AV--Large'' region is flatter
than that of the ``AV--Trans'' region.\footnote{%
    The comparison between the mid-IR extinction curve
    of the ``CO--Strong'' region with that of the ``CO--Weak''
    region is not very obvious. The overall shape of the
    ``CO--Strong'' curve appears flatter mainly because of
    the sudden rise of the IRAC 8$\mum$ band extinction.
   But we intend to think that the mid-IR extinction curve
   of the ``CO--Weak'' region is slightly flatter than
   that of the ``CO--Strong'' region because the extinction
   of the former at all the other three IRAC bands exceed
   that of the latter.
   }
This appears to support
the findings of McClure (2009) and Cambr\'esy et al.\ (2011)
who found that the mid-IR extinction curve becomes flatter
with the increasing of extinction.
However, the ``AV--Trans'' region does not seem to be more
heavily obscured than the ``CO--Strong'' region, even though
the former has a flatter mid-IR extinction curve than
the latter (see Figure~\ref{fig:irext}).
Furthermore, the ``Diffuse'' region which is the least densest
region among the five selected-regions,  has a mid-IR extinction
curve flatter than any of the other four regions.
Therefore, we conclude that we do not see a trend of
flattening the mid-IR extinction curve with increasing extinction.

The extinction in the IRAC 4.5$\mum$ band $A_{[4.5]}/A_\Ks$ is
significantly higher than that of the other three IRAC bands,
especially the ``Diffuse'' region has an exceedingly high
$A_{[4.5]}/A_\Ks$ value (see Figure~\ref{fig:irext} and
Table~\ref{tab:irext}). This 4.5$\mum$ extinction excess
will be discussed in \S\ref{sec:discussion}.

\begin{figure}
\centering
 \includegraphics[angle=0,width=6.5in]{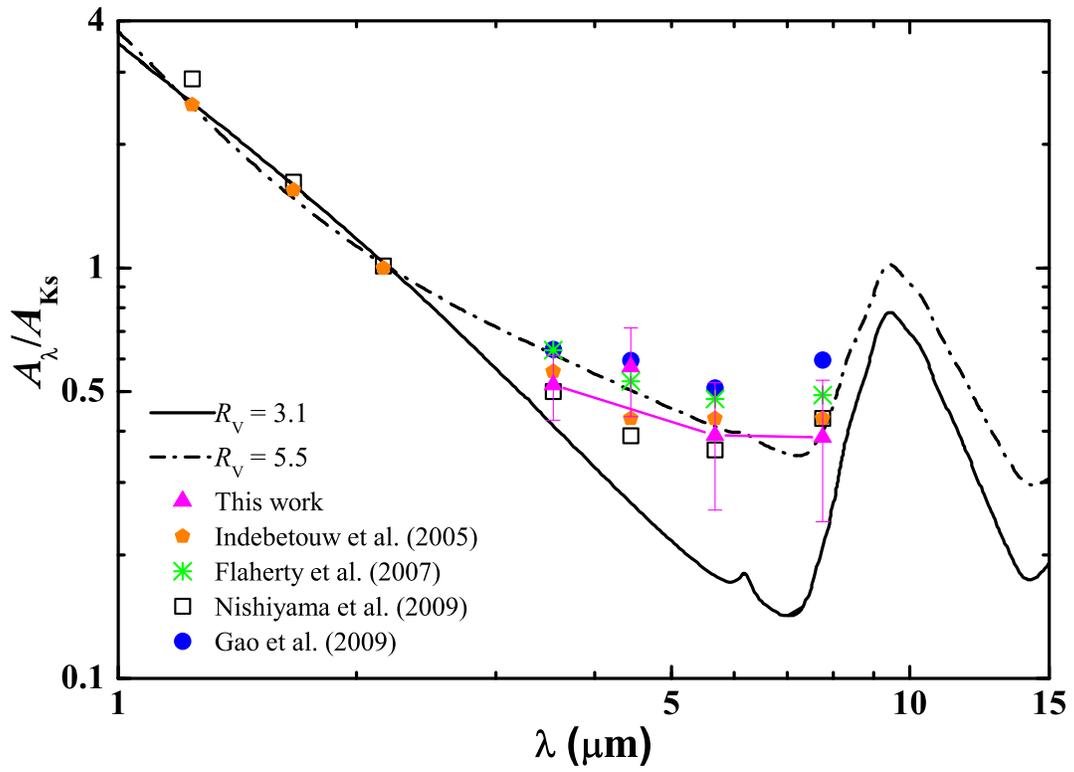}
\caption{\footnotesize
               \label{fig:irextmean}
               Comparison of the mid-IR extinction $A_\lambda/A_\Ks$
               averaged over the five selected regions
               with the previous determinations for other regions
               and the WD01 $\Rv=3.1, 5.5$ model extinction.
               The error bars indicate the standard deviation of five
               regions for each band.
               }
\end{figure}

Finally, we obtain the mean mid-IR extinction for the Coalsack nebula
by averaging over the five selected regions:
$A_{[3.6]}/A_\Ks \approx 0.519\pm0.093$,
$A_{[4.5]}/A_\Ks \approx 0.574\pm0.140$,
$A_{[5.8]}/A_\Ks \approx 0.391\pm0.134$,
and $A_{[8.0]}/A_\Ks \approx 0.387\pm0.146$.
The results are tabulated in Table~\ref{tab:irext} and
illustrated in Figure~\ref{fig:irextmean}.
Also shown in Figure~\ref{fig:irextmean} are the previous
determinations for other lines of sight
(e.g., Indebetouw et al.\ 2005, Flaherty et al.\ 2007,
Nishiyama et al.\ 2009, GJL09).
The mean mid-IR extinction derived here for the Coalsack nebula
is flat and roughly resembles the WD01 $\RV=5.5$ model extinction curve.
The 4.5$\mum$ extinction excess is also outstanding in the mean
mid-IR extinction curve.

\section{Discussion\label{sec:discussion}}
To examine the effects of dust size on the extinction,
we calculate the UV/optical/IR extinction from a mixture
of amorphous silicate and graphite.
We assume the standard ``MRN'' power-law size distribution
(Mathis et al.\ 1977) for both dust components:
$dn/da = A\,n_\HH\,a^{-3.5}$
for $a_{\min } < a < a_{\rm max}$,
where $a$ is the spherical radius of the dust
(we assume the dust to be spherical),
$a_{\rm min}=50\Angstrom$ is the lower cutoff of the dust size,
$a_{\rm max}$ is the upper cutoff,
$n_\HH$ is the number density of H nuclei,
and $A$ is constant
($A_{\rm sil}=10^{-25.11}\cm^{2.5}\,\HH^{-1}$ for the silicate component,
and $A_{\rm gra}=10^{-25.16}\cm^{2.5}\,\HH^{-1}$ for the graphitic
component; see  Draine \& Lee 1984).
For the upper cutoff of the dust size,
we take $a_{\rm max}=0.25, 0.5, 1, 2.5\mum$.
The $a_{\rm max}=0.25\mum$ case is known to closely
reproduce the Galactic mean extinction curve of $\RV=3.1$
(see Mathis et al.\ 1977, Draine \& Lee 1984).

\begin{figure}
\centering
\vspace{-0.6in}
\includegraphics[angle=0,width=4.2in]{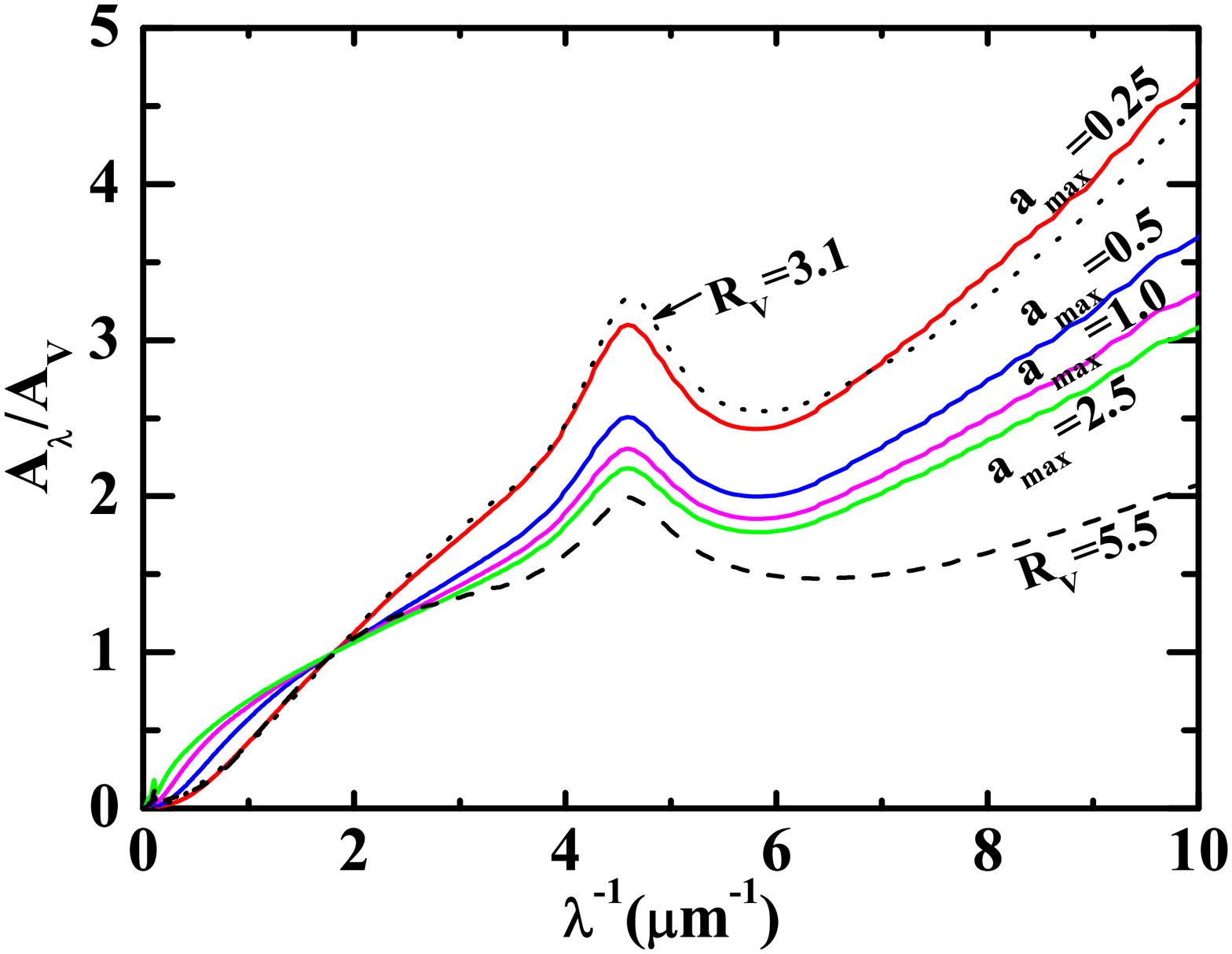}
\vspace{-0.2in}
\vspace{-0.15in}
\includegraphics[angle=0,width=4.2in]{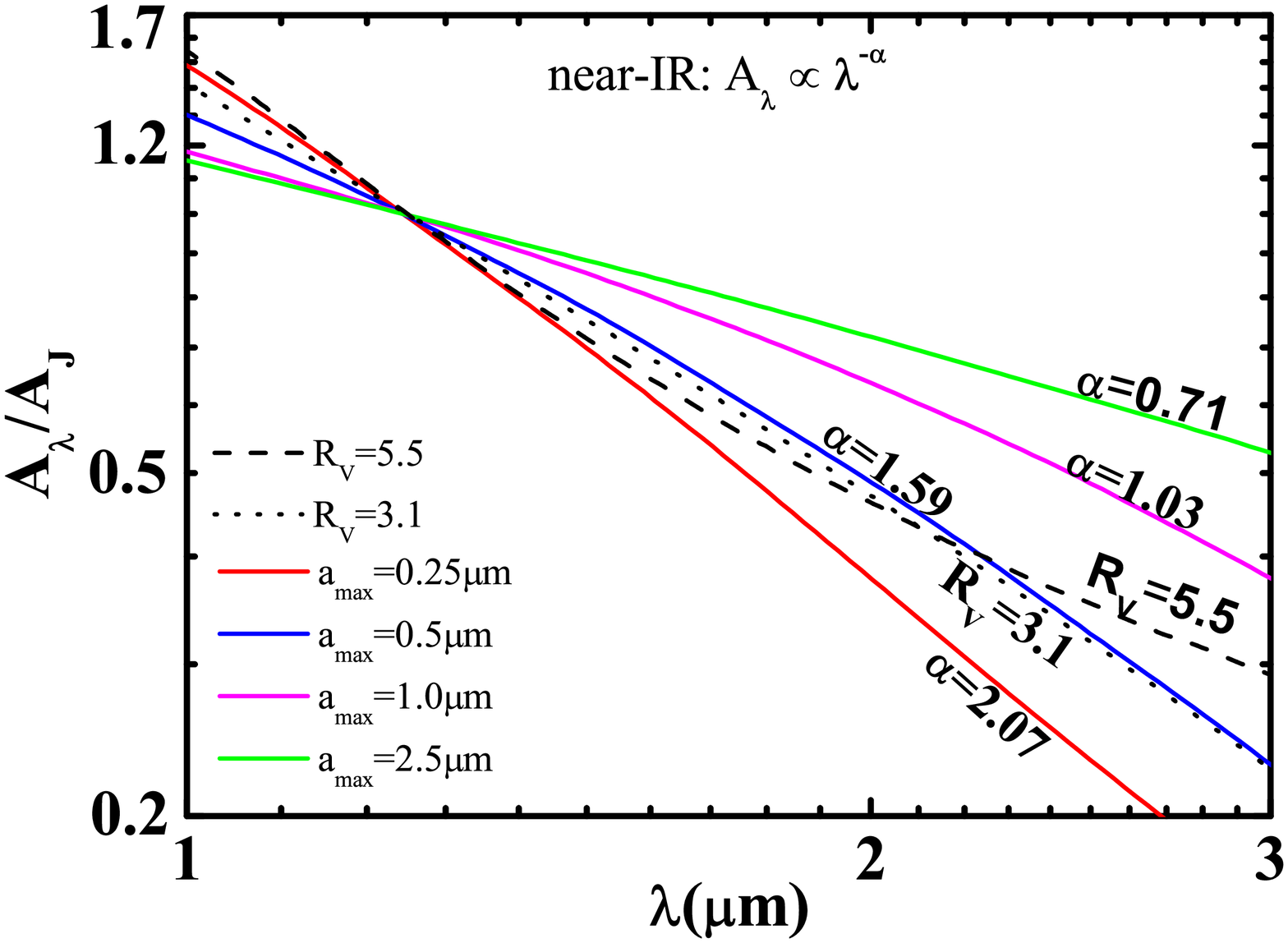}
\vspace{-0.2in}
\vspace{-0.2in}
\includegraphics[angle=0,width=4.2in]{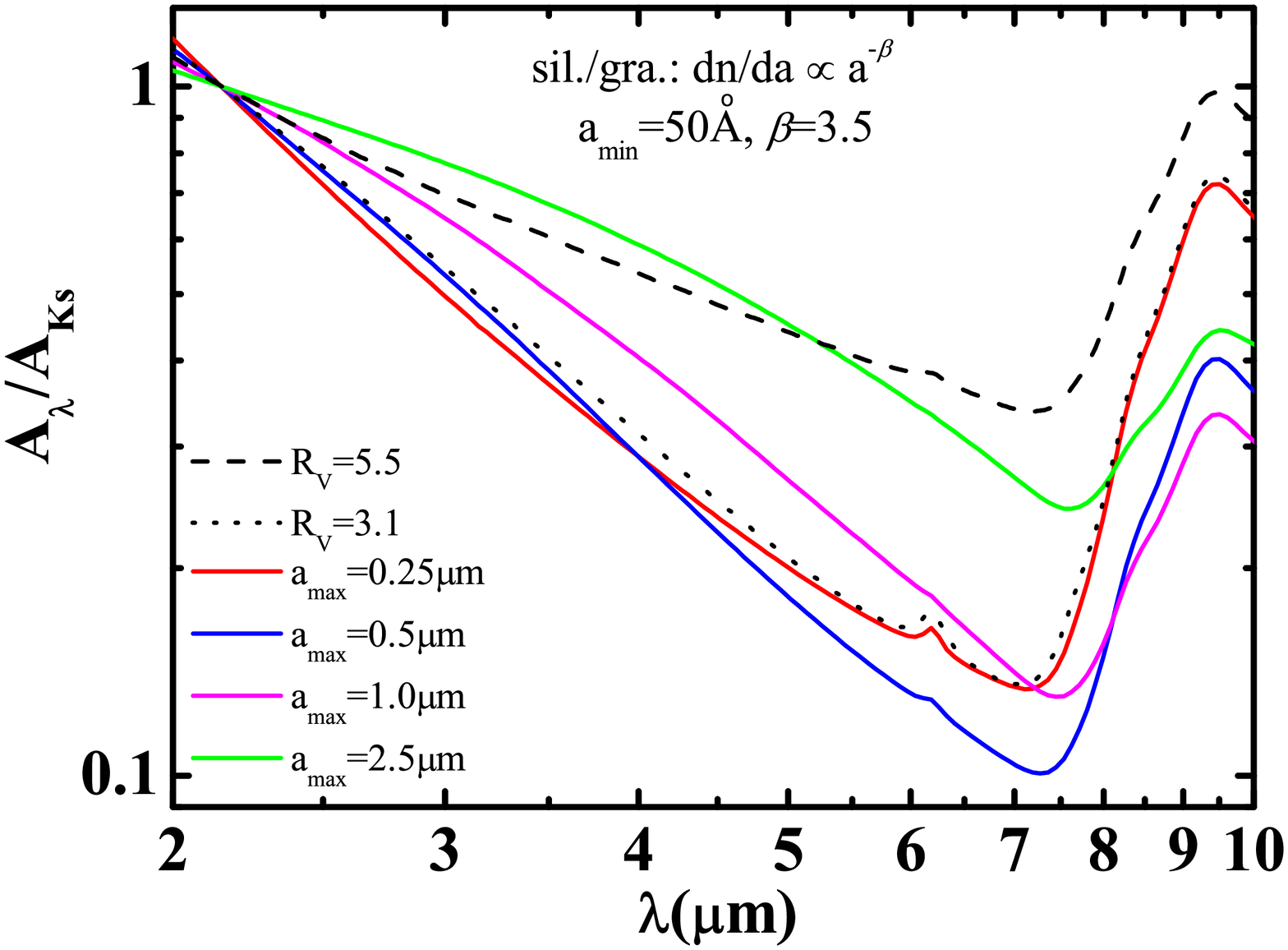}
\caption{\label{fig:amax}
               Upper panel (a): the UV/optical/near-IR extinction curves
               $A_\lambda/A_{\rm V}$
               calculated from a mixture of amorphous silicate and graphite
               with a power-law size distribution
               $dn/da\propto a^{-3.5}$ with $a_{\rm min}=50\Angstrom$
               and $a_{\rm max}=0.25\mum$ (red),
               $a_{\rm max}=0.5\mum$ (blue),
               $a_{\rm max}=1\mum$ (magenta), and
               $a_{\rm max}=2.5\mum$ (green).
               Also shown are the WD01 $\RV=3.1$ (dotted)
               and $\RV=5.5$ (dashed) model extinction curves.
               Middle panel (b): Same as (a) but for the near-IR
               extinction $A_\lambda/A_\JJ$.
               Lower panel (c): Same as (a) but for the mid-IR
               extinction $A_\lambda/A_\Ks$.
               }
\end{figure}

In Figure~\ref{fig:amax} we show
the UV/optical, near-, and mid-IR extinction
calculated from the dust mixture described above.
It is apparent that the UV extinction $A_\lambda/\AV$ flattens
with the increase of dust size. The near-IR extinction
$A_\lambda/A_\JJ$ also flattens with the increase of dust size:
if we approximate the near-IR extinction as a power-law
$A_\lambda\propto \lambda^{-\alpha}$, we would expect
a smaller $\alpha$ from a larger $a_{\rm max}$.
The variation of the mid-IR extinction $A_\lambda/A_\Ks$
with dust size is more complicated: unlike the UV and near-IR
extinction, the mid-IR extinction does not monotonically flatten
with the increase of $a_{\rm max}$, for example, $A_\lambda/A_\Ks$
steepens from $a_{\rm max}=0.25\mum$ to $a_{\rm max}=0.5\mum$,
but it flattens again when increasing $a_{\rm max}$ to 1$\mum$ or
larger (see Figure~\ref{fig:amax}).

The fact that the mid-IR extinction curve of  $a_{\rm max}=0.25\mum$
is even flatter than that of $a_{\rm max}=0.5\mum$ naturally
explains the regional variations of the mid-IR extinction of
the five regions selected in the Coalsack nebula:
the ``Diffuse'' region has the lowest density and the smallest grain
size and therefore the flattest mid-IR extinction;
the other regions are denser and therefore their grain sizes are
expected to be relatively larger and the resulting mid-IR extinction
curves are steeper. This trend is also seen in the ``CO--Strong''
region and the ``CO--Weak'' region: the former is denser and its
dust is larger and therefore its mid-IR extinction is relatively steeper.

However, it is puzzling that the mid-IR extinction of the
``AV--Large'' region is flatter than that of the ``AV--Trans''
region (and  that of the ``CO--Strong'' region). One would think
that the ``AV--Large'' region is denser than the ``AV--Trans''
region and its dust should be larger and therefore the mid-IR
extinction should be steeper.

The ``AV--Large'' region contains Globule 2,
the densest core in the Coalsack nebula.
\citet{Bok77} investigated this region
based on the  method of star counts
and found $\AV$ can reach $\simali$20\,mag toward
the center. The direction of the globule region has no known IRAS
sources \citep{Nyman89,Bourke95, Kato99}, protostellar objects
or young clusters \citep{Jones80}.
The lack of YSOs indicates that this region
has little star-forming activity.

\citet{Racca02} used a polytropic model to describe
the internal structure of this cloud, and argued that Globule 2
may be moderately unstable. \citet{Beuther11} investigated
the dynamical properties of the gas in Globule 2
with the APEX telescope through ${}^{13}$CO(2--1)
which suggests that this region exists dynamical action
(e.g., early infall activity). Hence, this cloud may be a young
molecular cloud complex and not yet centrally condensed.
The large $\AV$ may then be caused by the sightline crossing
long distance instead of large volume density of the dust.
Indeed, Smith et al.\ (2002) argued that the diffuse components
may contribute $\AV\sim5\magni$ to that of dense clouds in the
Coalsack globules.

\begin{figure}
\centering
 \includegraphics[angle=0,width=6.5in]{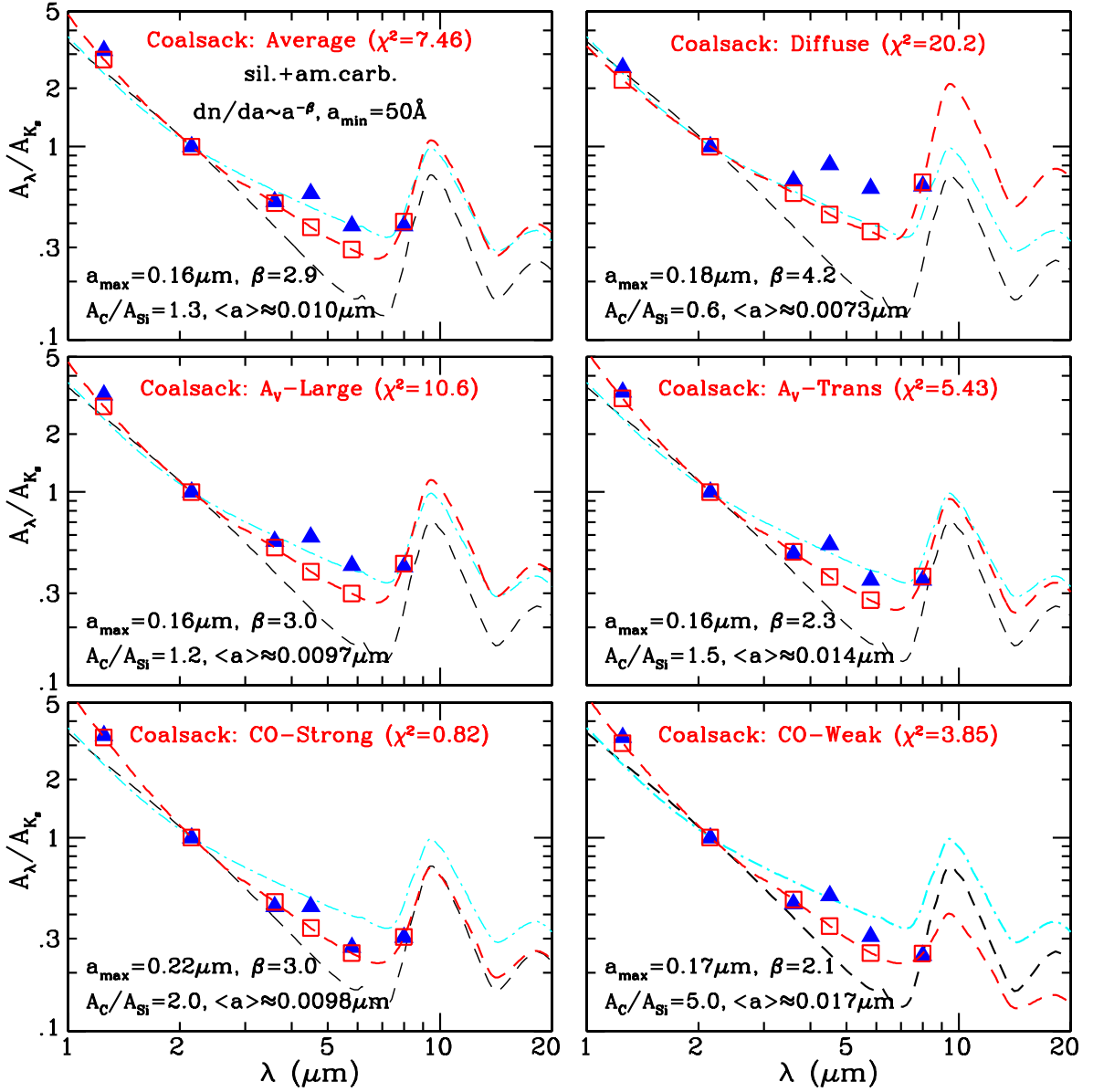}
\caption{\label{fig:irextmod}
              Comparison of the model extinction curve
              (red dashed line and red unfilled squares)
              with the IR extinction in the {\it 2MASS} and
              IRAC bands derived for the selected five regions of
              the Coalsack nebula as well as their mean values
              (blue filled triangles).
              The dust model consists of a mixture of amorphous
              silicate and amorphous carbon with an identical
              power-law size distribution for both dust components:
              $dn/da = A\,n_\HH\,a^{-\beta}$
              for $\amin <a< \amax$.
              With $\amin=50\Angstrom$ fixed,
              $\amax$, $\beta$, and $A_{\rm C}/A_{\rm Si}$
              (the volume ratio of amorphous carbon to silicate)
              are treated as free parameters.
              The goodness of the fit is measured by
              $\chi^2\equiv \sum \left\{\left(A_\lambda/A_\Ks\right)_{\rm
                mod} -  \left(A_\lambda/A_\Ks\right)_{\rm
         obs}\right\}^2/\left\{\Delta\left(A_\lambda/A_\Ks\right)\right\}^2$
              where $\left(A_\lambda/A_\Ks\right)_{\rm mod}$ is the
              model extinction, $\left(A_\lambda/A_\Ks\right)_{\rm
                obs}$ and $\Delta\left(A_\lambda/A_\Ks\right)_{\rm obs}$
              are the IR extinction and uncertainty at the {\it 2MASS} and
              IRAC bands derived for the selected regions.
              Also shown are the WD01 $\RV=3.1$ (black dashed line)
              and $\RV=5.5$ (cyan dot-dashed line)
              model extinction curves.
              }
\end{figure}

We stress that the above findings of steeper mid-IR extinction curves
in denser regions are true only when the dust grains are small:
as illustrated in Figure~\ref{fig:amax}, the mid-IR extinction deepens
from $\amax=0.25\mum$ to $\amax=0.5\mum$, while it flattens
again from $\amax=0.5\mum$ to $\amax=1\mum$.
To quantitatively examine this, we model the IR extinction curves
in the {\it 2MASS} and IRAC bands derived for the selected five
regions and their mean values in terms of a mixture of amorphous
silicate and amorphous carbon.\footnote{%
  We have considered graphite. But the mid-IR extinction is
  better fitted with amorphous carbon. This is because disordered
  carbon materials like amorphous carbon have many IR active
  modes so that they better fit the flat mid-IR extinction.
  }
Both dust components are assumed to have the same size distribution
$dn/da = A\,n_\HH\,a^{-\beta}$ for $\amin <a< \amax$
with $\amin=50\Angstrom$. The free parameters are
$\amax$, $\beta$, and the volume ratio of carbon dust to
silicate dust $A_{\rm C}/A_{\rm Si}$.
In Figure~\ref{fig:irextmod} we show the best-fit results
for the selected five regions and their mean values.
The model fits the derived extinction reasonably well
except the IRAC [4.5] and [5.8] bands.
The mean dust size
$\langle a \rangle =
\int_{\amin}^{\amax} a\,\left(dn/da\right)\,da/\int_{\amin}^{\amax}
\left(dn/da\right)\,da$
is consistent with
the flatness of the mid-IR extinction:
the smaller $\langle a \rangle$ is,
the flatter is the mid-IR extinction
(e.g., the ``AV--Large'' region with
$\langle a \rangle \approx 0.0097\mum$
has a flatter mid-IR extinction
curve than that of the ``AV--Trans'' region
for which $\langle a \rangle \approx 0.014\mum$).
The mean dust size $\langle a \rangle$
is the smallest for the ``Diffuse'' region
which has the flattest mid-IR extinction.
The fact that the mid-IR extinction curve steepens with
the increase of $\langle a \rangle$ indicates that the dust
in the five regions in the Coalsack nebula is small:
the dust has not yet grown sufficiently to the level that
the mid-IR extinction flattens with the increase of
$\langle a \rangle$ (see Figure~\ref{fig:amax}).
This suggests that the chemical evolution of
the Coalsack nebula may be at too early a stage
for the dust to significantly grow.

The extinction in the IRAC 4.5$\mum$ band,
$A_{[4.5]}/A_\Ks$, is significantly higher than that
of the other three IRAC bands
(see Figure~\ref{fig:irext}) and stands
well above the model curve for all regions
(see Figure~\ref{fig:irextmod}).
The 4.5$\mum$ extinction ``excess'' is particularly
prominent for the ``Diffuse'' region.
As clearly seen in Figure~\ref{fig:irextmean}
and Table~\ref{tab:irext}, the mean extinction of
$A_{[4.5]}/A_\Ks\approx0.574$ exceeds that of
most of the other regions previously determined by others,
and is comparable to the Galactic plane average of
$A_{[4.5]}/A_\Ks\approx0.57$ derived by GJL09.
The detection of the 4.5$\mum$ extinction excess has also
been reported in the LMC (see Gao et al.\ 2013).

Could the 4.5$\mum$ extinction ``excess'' arise from
interstellar CO and CO$_2$ ices?
CO ice has an absorption feature at 4.67$\mum$
due to the C--O stretch,
while CO$_2$ ice has an absorption feature at 4.27$\mum$.
These absorption features could contribute to the IRAC 4.5$\mum$
extinction. If CO and CO$_2$ ices are present,
one would expect H$_2$O ice to be present as well,
with an abundance usually much higher than that of
CO and CO$_2$ ices
(Whittet 2003, Gibb et al.\ 2004, Boogert et al.\ 2011).
H$_2$O ice has two strong absorption bands:
the O--H stretching feature at 3.05$\mum$
and the H--O--H bending feature at 6.02$\mum$.

\begin{figure}
\centering
 \includegraphics[angle=0,width=6.5in]{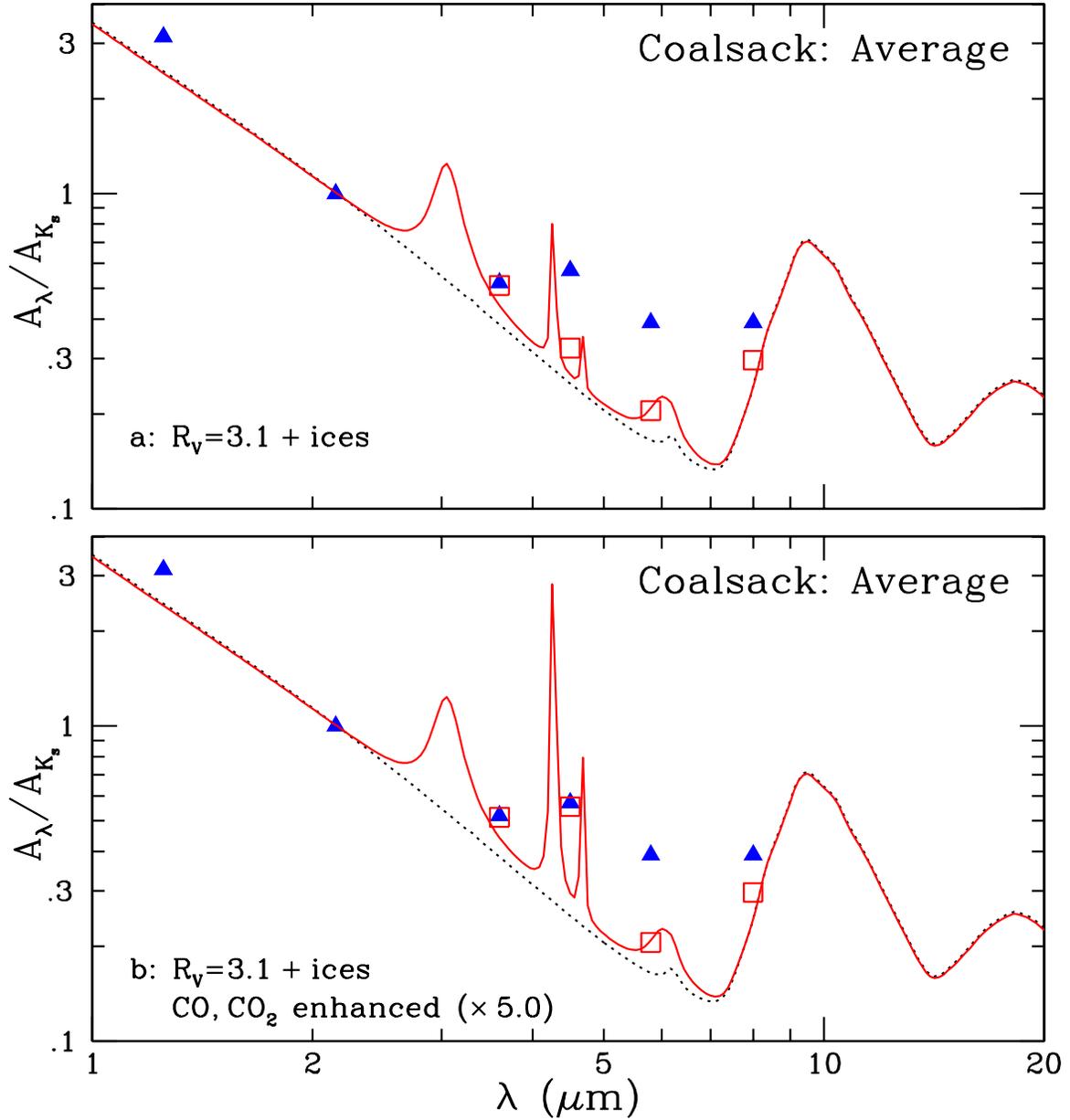}
\caption{\footnotesize
         \label{fig:ice.ext}
         Comparison of the average extinction
         derived for the five selected regions
         of the Coalsack nebula (blue filled triangles)
         with the $\Rv=3.1$ model extinction curve
         (black dotted line)
         combined with the 3.05 and 6.02$\mum$ absorption
         bands of H$_2$O ice,
         the 4.27$\mum$ band of CO$_2$ ice,
         and the 4.67$\mum$ band of CO ice
         (red solid line).
         Open red squares are
         the $\Rv=3.1$ model curve
         plus the ice absorption bands convolved
         with the {\it Spitzer}/IRAC filter functions.
         The little bump at 6.2$\mum$ in the $R_{\rm V}=3.1$ model curve (black dotted line) is due to the C--C stretching absorption feature of polycyclic aromatic hydrocarbon (PAH) material (see Li \& Draine 2001).
         Upper panel: the abundances of
         CO and CO$_2$ ices are taken to be that
         of typical dense clouds
         (i.e., ${\rm CO/H_2O =0.25}$,
          ${\rm CO_2/H_2O =0.21}$).
         Bottom panel: the abundances of
         CO and CO$_2$ ices are enhanced
         by a factor of five relative to
         their typical values in dense clouds.
         }
\end{figure}

To examine whether the ice (particularly CO$_2$
and, to a less degree, CO) absorption bands could account for
the 4.5$\mum$ excess extinction, we approximate the 3.05, 4.27, 4.67
and 6.02$\mum$ bands respectively from H$_2$O, CO$_2$, CO,
and H$_2$O ices as four Drude profiles. The extinction due to these
four ice bands is
\begin{equation}\label{eq:ice1}
A_{\lambda}({\rm ice}) = 1.086\,N_{\rm H_{2}O} \sum\limits_{j=1}^4
\frac{A_j \times \left[{\rm X}_j/{\rm H_{2}O}\right] \times E_j
\times 2\gamma_j/\pi}
{\left(\lambda -\lambda_j^2/\lambda\right)^2+\gamma_j^2}
\end{equation}
where
$\lambda_j$, $\gamma_j$, $A_j$ are the peak wavelength,
FWHM, and strength of the $j$-th ice absorption band, respectively
(see Table~\ref{tab:ice});\footnote{%
   The FWHM $\gamma_j$ and band strength $A_j$
   values tabulated in Table~\ref{tab:ice} are in units of
   cm$^{-1}$ and cm\,molecule$^{-1}$, respectively.
   When using eq.\,\ref{eq:ice1},
   they are converted so that they are in units of
   $\mum$ and  cm$^3$\,molecule$^{-1}$.
    }
$N_{\rm H_2O}$ is the H$_2$O ice column density;
${\rm X}_j/{\rm H_2O}$ is the abundance of
the ice species ${\rm X}_j$ relative to H$_2$O ice
in typical dense clouds;
$E_j$ is the ``enhancement'' factor for species ${\rm X}_j$
(i.e., ${\rm X}_j/{\rm H_2O}$ is increased by
a factor of $E_j$ in order for the ice absorption bands
to account for the 4.5$\mum$ excess extinction).
Finally, we add $\left(A_\lambda/A_\Ks\right)_{\rm ice}$,
the ice extinction $A_\lambda({\rm ice})$ normalized to
 $A_\Ks({\rm ice})$,
the ice extinction at the $\Ks$ band,
to
$\left(A_\lambda/A_{\Ks}\right)_{\rm WD}$,
the WD01 model extinction curve (with $\Rv=3.1$),\footnote{%
  Ideally, we should add the ice extinction to the best-fit
  model extinction curves (see Figure~\ref{fig:irextmod}).
  However, the models have already closely fitted
  the extinction at the 3.6$\mum$ IRAC band and therefore
  no ice can be allowed: with the addition of any ice extinction,
  the resulting 3.6$\mum$ extinction would exceed
  the observationally-derived value.
  This is also true for the $\Rv=5.5$ WD01 model
  extinction curve. Therefore, we only consider
  the $\Rv=3.1$ model curve.
  }
\begin{equation}\label{eq:ice2}
A_\lambda/A_\Ks = \left(1-f_{\rm ice}\right)\times
                  \left(A_\lambda/A_{\Ks}\right)_{\rm WD}
+ f_{\rm ice}\times\left(A_\lambda/A_\Ks\right)_{\rm ice} ~~,
\end{equation}
where $f_{\rm ice}$ is the fractional amount of extinction
at the $\Ks$ band contributed by ices.
We then convolve $A_\lambda/A_\Ks$
with the {\it Spitzer}/IRAC
filter response functions
and, for demonstrative purposes,
compare with the average mid-IR extinction
derived for the Coalsack nebula
(see Figure~\ref{fig:ice.ext}).

The H$_2$O ice quantity (i.e., $f_{\rm ice}$)
is constrained not to exceed the IRAC 3.6$\mum$ extinction.
With ${\rm CO/H_2O \approx 0.25}$
and ${\rm CO_2/H_2O \approx 0.21}$,
typical for quiescent dense clouds
(Whittet 2003, Gibb et al.\ 2004, Boogert et al.\ 2011),
as shown in Figure~\ref{fig:ice.ext}a,
CO and CO$_2$ are not capable of accounting for
the 4.5$\mum$ excess extinction.
If we are forced to attribute the 4.5$\mum$ excess extinction
to CO and CO$_2$, we will have too much H$_2$O ice
and the resulting extinction at 3.6$\mum$ would be too high.
If we fix the H$_2$O ice abundance at what is required by
the 3.6$\mum$ extinction, in order for CO and CO$_2$ to explain
the 4.5$\mum$ excess extinction, we have to enhance their abundances
(relative to their typical abundances in dense clouds)
by a factor of
$E_{\rm CO}=E_{\rm CO_2}\approx 5$ for $\Rv=3.1$
(see Figure~\ref{fig:ice.ext}b).
The required CO and CO$_2$ abundances are
unrealistically too high:
${\rm CO/H_2O \approx 1.3}$
and ${\rm CO_2/H_2O \approx 1.1}$.
Also, the model still could not account for
the flat extinction at 5.8$\mum$ and 8.0$\mum$
(see Figure~\ref{fig:ice.ext}b).

The 3.1$\mum$ H$_2$O ice absorption band has been detected
in the Coalsack nebula. \citet{Smith02} observed the lines of sight
toward eight field stars lying behind three of the densest globules
in the Coalsack nebula (i.e., Globules 1, 2, and 3)
with the CTIO 4\,m telescope and its IR spectrometer.
They reported the detection of strong ice absorption
in two sources (SS1-2 in Globule 1 and D7 in Globule 2),
moderate or weak ice absorption in five sources
(D4, D11, E9, and E18 in Globule 2, and SS3-1 in Globule 3),
and very weak or no ice absorption in one source
(E2 in Globule 2).
The sightlines toward these stars are ideal for investigating
the regional variations of the mid-IR extinction and particularly
for examining the nature of the IRAC 4.5$\mum$ extinction excess.
The optical depth of the 3.1$\mum$ ice absorption feature
$\tau_{\rm ice}$ probes well whether a region is dense or diffuse:
it is only seen in dense clouds and it is generally true that
the higher $\tau_{\rm ice}$ is, the denser is the cloud.
If the IRAC 4.5$\mum$ extinction excess is caused by CO and CO$_2$
ices, it should be more prominent in sightlines with strong ice
absorption (e.g., D7 in Globule 2) and much less prominent or
even disappear in sightlines with very weak or no ice absorption
(e.g., E2 in Globule 2)

We use the ``color-excess'' method (see \S3.1) and take RG stars
as tracers to derive the mid-IR extinction in the IRAC bands for
the six Globule 2 sightlines studied by Smith et al.\ (2002).
In order to have enough sources (for each sightline) which
can be used to derive extinction statistically, we consider
an area of $4.8'\times4.2'$ centered around each given star.
As D4 and D11 are very close and their ice absorption strengths
are comparable, we combine them into one region
(hereafter ``D4/D11'').
The region that contains E9 is not considered since it
overlaps parts of E2 and E18 and is difficult to
separate it from them.
Therefore, we have four regions for IR extinction studies:
E2 -- very weak ice absorption
($\tau_{\rm ice}\approx 0.01\pm0.003$),
E18 -- moderate or weak ice absorption
($\tau_{\rm ice}\approx 0.07\pm0.004$),
D4/D11 -- moderate or weak ice absorption
($\tau_{\rm ice}\approx 0.08\pm0.004$), and
D7 -- strong ice absorption
($\tau_{\rm ice}\approx 0.34\pm0.007$).
The parameters (including the number of RG stars)
of these regions are tabulated in Table~\ref{tab:icesources}.
In Table~\ref{tab:iceirext} we tabulate the near-IR extinction
power exponent $\alpha$ derived from the {\it 2MASS} photometry
and the extinction ratio $A_{\lambda}/A_\Ks$ in the four IRAC bands.

\begin{figure}
\centering
 \includegraphics[angle=0,width=6.5in]{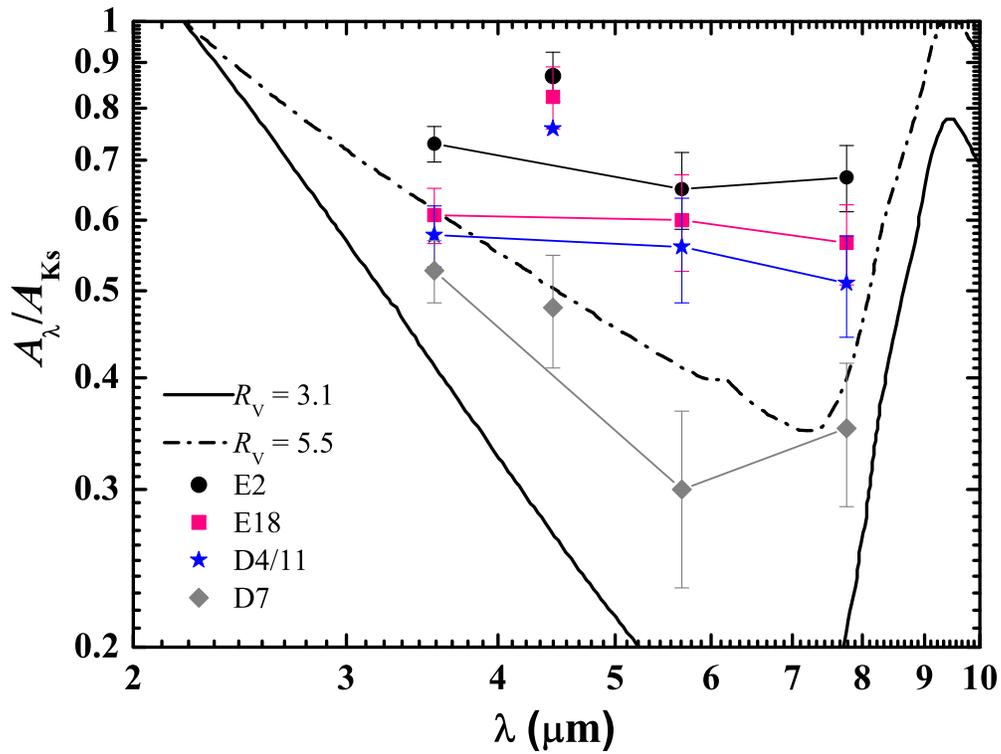}
\caption{\label{fig:iceregions}
               $A_\lambda/A_\Ks$ in the four IRAC bands
               of the four representative regions in the Coalsack
               nebula Globule 2 spanning a range of ice absorption
               strengths: D7 -- strong ice absorption,
               D4/D11 and E18 -- moderate or weak ice absorption,
               and E2 -- very weak or no ice absorption.
               For comparison, the WD01 $\Rv=3.1, 5.5$ model
               extinction curves are also shown.
               }
\end{figure}

The results are illustrated in Figure~\ref{fig:iceregions}.
It is apparent that $A_{\lambda}/A_\Ks$ in the IRAC bands increases
with the decrease of $\tau_{\rm ice}$. This is consistent with our
results for the ``Diffuse'', ``AV--Large'', ``AV--Trans'',
``CO--Strong'' and ``CO--Weak'' regions: the mid-IR extinction
curve flattens in less dense regions (provided that the dust is
small), suggesting that the dust in the Coalsack nebula
has not grown sufficiently even in the dense cloud sightline.
Smith et al.\ (2002) found that in the Coalsack nebula,
above a threshold of $\AV\sim 7.6\magni$,
the 3.1$\mum$ H$_2$O absorption optical depth $\tau_{\rm ice}$
correlates with the visual extinction $\AV$:
$\tau_{\rm ice} \approx 0.059\,\left(\AV-7.6\right)$.
Such a correlation is also seen in the Taurus dark cloud
(a quiescent molecular cloud free from embedded young
high mass stars) but with a smaller $\AV$ threshold:
$\tau_{\rm ice} \approx 0.059\,\left(\AV-2.6\right)$
(Whittet et al.\ 1988b, Smith et al.\ 1993, Whittet et al.\ 2001).
Smith et al.\ (2002) argued that there may exist a diffuse component
contributing $\AV\sim5\magni$ to the extinction of the dense globules.
Therefore, even the total visual extinction $\AV$ is large toward
a sightline, it may not necessarily be very dense (it might just be
a pile-up of a long column of diffuse material). This also explains
why the ``AV--Large'' region has a flatter mid-IR extinction curve
than the ``AV--Trans'' region.
This also explains why the dust in the Coalsack nebula has not grown
effectively, just like that the ice mantles have not been effectively built
up on grains.

We note that IRAC 4.5$\mum$ extinction excess is prominent in
E2, the sightline with very weak or no ice absorption. On the other
hand, although the 4.5$\mum$ extinction excess is also seen in
D7 -- the sightline with strong ice absorption, it is much weaker.
No detection of the 4.27$\mum$ CO$_2$ ice feature
or the 4.67$\mum$ CO ice feature was reported.
Therefore, the 4.5$\mum$ extinction excess cannot be attributed
to CO and CO$_2$ ices.

The origin of the 4.5$\mum$ excess extinction remains unknown.
Some red giants have circumstellar envelopes rich in CO gas
which absorbs at 4.6$\mum$\citep{Bernat81}.
With red giants as a tracer of the mid-IR extinction,
the 4.6$\mum$ absorption feature of their CO gas
could result in an overestimation of the IRAC 4.5$\mum$ extinction.
It is worth exploring whether the CO gas absorption of red giants
could account for the 4.5$\mum$ excess extinction.

\section{Summary}

Using data from the near-IR {\it 2MASS} Survey
and the mid-IR \textit{Spitzer}/GLIPMSE Legacy Program,
we have derived the mid-IR extinction curves of
five regions in the starless Coalsack nebula,
spanning a range of interstellar environments
from diffuse through translucent to dense clouds.
The (relative) extinction $A_{\lambda}/A_\Ks$
in the four \textit{Spitzer}/IRAC bands ([3.6], [4.5], [5.8],
[8.0]$\mum$) are determined using the color-excess method
and taking red giants as tracers. It is found that:
\begin{enumerate}
\item The mid-IR extinction curve exhibits appreciable
          regional variations. The extinction ratios
          $A_\lambda/A_\Ks$ in the IRAC bands
          are higher in diffuse regions than in denser regions.
          These variations can be explained in terms of
          dust size effects: the dust in dense regions is
          larger than that in diffuse regions, provided that
          the dust is small (i.e., it has not yet grown effectively
          in the Coalsack nebula).
\item The mean mid-IR extinction curve,
          with $A_{[3.6]}/A_\Ks\approx0.519$,
          $A_{[4.5]}/A_\Ks\approx0.574$,
          $A_{[5.8]}/A_\Ks\approx0.391$, and
          $A_{[8.0]}/A_\Ks\approx0.387$,
          is flat and consistent with previous results for various
          regions and the WD01 model extinction curve of $\RV=5.5$.
\item The extinction in the IRAC 4.5$\mum$ band
          is much higher than that of the other three IRAC bands.
          It cannot be explained in terms of
          the 4.27$\mum$ absorption band of $\rm CO_2$ ice
          and the 4.67$\mum$ absorption band of CO ice.
          It may be caused by the 4.6$\mum$ absorption feature
          of CO gas in the circumstellar envelopes of some red giants.
\item The mid-IR extinction in the four IRAC bands have also
          been derived for four regions in the Coalsack Globule 2
          which respectively exhibit strong ice absorption,
          moderate or weak ice absorption,
          and very weak or no ice absorption.
          The derived mid-IR extinction curves are all flat,
           with $A_{\lambda}/A_\Ks$ (in the IRAC bands) increasing
           with the decrease of the 3.1$\mum$ H$_2$O ice absorption
           optical depth $\tau_{\rm ice}$.
\end{enumerate}

\acknowledgments{We thank A.C.A.~Boogert and E.~Gibb
for helpful discussions.
We thank the anonymous referee for
helpful suggestions.
This project is supported in part through China's grants
NSFC No.\, 11173007 and 11173019.
AL is supported in part by NSF/AST 1109039, NNX13AE63G,
and the University of Missouri Research Board.
We also thank the John Templeton Foundation in conjunction
with National Astronomical Observatories, Chinese Academy of Sciences.}

\clearpage

\clearpage

\clearpage

\clearpage

\clearpage

\clearpage

\clearpage

\clearpage

\clearpage

\clearpage

\clearpage

\clearpage

\clearpage

\clearpage

\begin{table}
\begin{center}
\caption{\label{tab:5regions}
               Five regions in the Coalsack nebula
               selected for the IR extinction studies.
                }
\vspace{0.2in}
\begin{tabular}{lcccc}
\hline \hline
Selected &  $l$   &   $b$ & Area    & Number of \\
Region  &  (deg)   &  (deg) & Covered   & Sources\\
\hline
Diffuse       &    306.45     &    -0.85     & $18'\times6'$  & 245  \\
AV--Large     &    300.75     &    -0.95     & $30'\times12'$ & 951  \\
AV--Trans     &    300.75     &    -0.35     & $30'\times12'$ & 800  \\
CO--Strong    &    303.40     &    -0.30     & $36'\times24'$ & 2769 \\
CO--Weak      &    303.38     &    -0.96     & $30'\times10.5'$ & 657 \\
\hline
\end{tabular}
\end{center}
\end{table}

\begin{table}
\begin{center}
\caption{\label{tab:colorexcess}
               The {\it 2MASS} and {\it Spitzer}/IRAC color excess ratios
               for the five selected regions and their average values.
               }
\vspace{0.2in}
\begin{tabular}{lccccc}
\hline \hline
Region  &\scriptsize$E_{\J-\HH}/E_{\HH-\Ks}$
&\scriptsize$E_{\Ks-[3.6]}/E_{\J-\Ks}$
&\scriptsize$E_{\Ks-[4.5]}/E_{\J-\Ks}$
&\scriptsize$E_{\Ks-[5.8]}/E_{\J-\Ks}$
&\scriptsize$E_{\Ks-[8.0]}/E_{\J-\Ks}$\\
\hline
Diffuse    &1.680$\pm$0.062 &0.216$\pm$0.018 &0.125$\pm$0.028 &0.251$\pm$0.024 &0.241$\pm$0.023\\
AV--Large  &1.876$\pm$0.019 &0.203$\pm$0.004 &0.188$\pm$0.007 &0.265$\pm$0.006 &0.269$\pm$0.006\\
AV--Trans  &1.908$\pm$0.024 &0.226$\pm$0.005 &0.203$\pm$0.008 &0.283$\pm$0.008 &0.283$\pm$0.008\\
CO--Strong &1.928$\pm$0.009 &0.237$\pm$0.002 &0.236$\pm$0.003
&0.308$\pm$0.003 &0.294$\pm$0.003\\
CO--Weak   &1.909$\pm$0.025 &0.237$\pm$0.006 &0.218$\pm$0.011 &0.301$\pm$0.009 &0.328$\pm$0.010\\
\hline
Mean       &1.860$\pm$0.010 &0.224$\pm$0.015 &0.194$\pm$0.043 &0.282$\pm$0.024 &0.283$\pm$0.032\\
\hline
\end{tabular}
\end{center}
\end{table}

\begin{table}
{\small
\begin{center}
\caption{\label{tab:irext}
               IR extinction (relative to the {\it 2MASS} $\Ks$ band
               extinction) of the five selected regions in the
               Coalsack nebula.
               }
\vspace{0.2in}
\begin{tabular}{lcccccc}
\hline \hline
Region     &$\alpha$      &$A_\J/A_\Ks$  &$A_{[3.6]}/A_\Ks$&$A_{[4.5]}/A_\Ks$&$A_{[5.8]}/A_\Ks$&$A_{[8.0]}/A_\Ks$ \\
\hline
Diffuse    &1.73$\pm$0.01 &2.56$\pm$0.01 &0.665$\pm$0.028 &0.806$\pm$0.043 &0.609$\pm$0.038 &0.625$\pm$0.035 \\
AV--Large  &2.14$\pm$0.01 &3.19$\pm$0.01 &0.556$\pm$0.009 &0.587$\pm$0.016 &0.418$\pm$0.014 &0.410$\pm$0.013 \\
AV--Trans  &2.20$\pm$0.01 &3.28$\pm$0.01 &0.481$\pm$0.011 &0.534$\pm$0.019 &0.351$\pm$0.018 &0.351$\pm$0.018 \\
CO--Strong &2.24$\pm$0.01 &3.37$\pm$0.01 &0.439$\pm$0.004
&0.441$\pm$0.008 &0.270$\pm$0.007 &0.304$\pm$0.007 \\
CO--Weak   &2.20$\pm$0.01 &3.28$\pm$0.01 &0.456$\pm$0.013 &0.500$\pm$0.025 &0.308$\pm$0.021 &0.246$\pm$0.023 \\
\hline
Mean       &2.10$\pm$0.21 &3.14$\pm$0.33 &0.519$\pm$0.093 &0.574$\pm$0.140 &0.391$\pm$0.134 &0.387$\pm$0.146 \\
\hline
\end{tabular}
\end{center}
}
\end{table}

\clearpage

\begin{table}  
\small
\begin{center}
\caption{%
         \label{tab:ice}
         Parameters for the four ice bands of H$_2$O, CO and
         CO$_2$ which may contribute to the mid-IR extinction
         in the Spitzer/IRAC bands.
         }
\vspace{0.2in}
\begin{tabular}{lccccccr}
\hline\hline
Band & $\lambda_j$ & Ice Species & FWHM\tablenotemark{a}
        & Band Strength\tablenotemark{b} & Abundance\tablenotemark{c}
        & Enhancement\tablenotemark{d} & $f_{\rm ice}$\tablenotemark{e}\\
$j$     & ($\mu$m) & ${\rm X}_j$ & $\gamma_j$ (cm$^{-1}$)
        & $A_j$ (cm\,mol$^{-1}$)  & ${\rm X}_j/{\rm H_2O}$
        & $E_j$ ($\Rv=3.1$) & \\
\hline
1 & 3.05 & H$_{2}$O  & 335
  & $2.0\times10^{-16}$  & 1
  & 1 & 0.015\\
2 & 4.27 &CO$_{2}$ & 18
  & $7.6\times10^{-17}$  & 0.21
  & 5 
  & 0.015\\ 
3 & 4.67 &CO   & 9.7
  & $1.1\times10^{-17}$  & 0.25
  & 5 & 0.015\\
4 & 6.02 &H$_{2}$O & 160
  & $8.4\times10^{-16}$  & 1
  & 1 & 0.015\\
\hline
\end{tabular}
\tablenotetext{a}{Gibb et al.\ (2004)}
\tablenotetext{b}{Gerakines et al.\ (1995)}
\tablenotetext{c}{Whittet (2003)}
\tablenotetext{d}{$E_j$ is the ``enhancement'' factor
                  in the sense that in order for CO and
                  CO$_2$ ices to account for the 4.5$\mum$
                  excess extinction, the CO and CO$_2$
                  abunadnces need to be ``enhanced''
                  by a factor of $E_j$ (i.e., $E_{\rm CO}$
                  and $E_{\rm CO_2}$) relative to their
                  abundances in typical dense clouds.}
\tablenotetext{e}{$f_{\rm ice}$ is the fractional amount
                  of extinction at the $\Ks$ band
                  contributed by ices.}
\end{center}
\normalsize
\end{table}


\begin{table}
\begin{center}
\caption{\label{tab:icesources}
               Four representative lines of sight toward E2, E18,
               D4/D11, and D7 in the Coalsack Globule 2 which
               span a range of ice absorption strengths.
               }
\vspace{0.2in}
\begin{tabular}{lcccccc}
\hline \hline
Name   & Water Ice      &$\tau_{\rm ice}$&  $l$    &   $b$   & Area      & \small Number of \\
       & Absorption     &                &  (deg)  &  (deg)  & Covered   & \small Sources\\
\hline
E2     & Very Weak      & 0.01$\pm$0.03  & 300.65     &    -0.835    & $4.8'\times4.2'$ & 43  \\
E18    & Moderate/Weak  & 0.07$\pm$0.03  & 300.76     &    -0.775    & $4.8'\times4.2'$ & 61  \\
D4/D11 & Moderate/Weak  & 0.08$\pm$0.04  & 300.69     &    -0.905    & $4.8'\times4.2'$ & 50 \\
D7     &  Strong        & 0.34$\pm$0.07  & 300.71     &    -0.975    & $4.8'\times4.2'$ & 46 \\
\hline
\end{tabular}
\end{center}
\end{table}

\begin{table}
\small
\begin{center}
\caption{\label{tab:iceirext}
               IR extinction (relative to the {\it 2MASS} $\Ks$ band
               extinction) of the regions in the
               Coalsack Globule 2
               which span a range of ice absorption strengths.
               }
\vspace{0.2in}
\begin{tabular}{lcccccc}
\hline \hline
Name   & Ice Absorption&$\alpha$       &$A_{[3.6]}/A_\Ks$&$A_{[4.5]}/A_\Ks$&$A_{[5.8]}/A_\Ks$&$A_{[8.0]}/A_\Ks$ \\
\hline
E2     & Very Weak      &1.72$\pm$0.01   &0.730$\pm$0.033 &0.869$\pm$0.055 &0.650$\pm$0.064 &0.670$\pm$0.057 \\
E18    & Moderate/Weak  &1.91$\pm$0.01   &0.608$\pm$0.043 &0.824$\pm$0.066 &0.600$\pm$0.074 &0.566$\pm$0.058 \\
D4/D11 & Moderate/Weak  &2.14$\pm$0.01   &0.577$\pm$0.045 &0.759$\pm$0.077 &0.560$\pm$0.075 &0.510$\pm$0.066 \\
D7     & Strong         &2.16$\pm$0.01   &0.527$\pm$0.042 &0.479$\pm$0.069 &0.300$\pm$0.067 &0.351$\pm$0.064 \\
\hline
\end{tabular}
\end{center}
\end{table}

\end{document}